\documentclass{emulateapj}
\shorttitle{Optical/Near-IR Selected Galaxies at $Z\sim 2$}
\shortauthors{Reddy et al.}
\slugcomment{Received 2005 April 21; Accepted 2005 July 7}

\begin{document}
\newcommand{\ebmv}{E(B-V)}
\newcommand{\sfr}{{\rm M}_{\odot} ~ {\rm yr}^{-1}}
\newcommand{\zmk}{(z-K)_{\rm AB}}
\newcommand{\jmk}{J-K_{\rm s}}
\newcommand{\rmk}{{\cal R}-\ks}
\newcommand{\ugr}{{\rm BX/BM}}
\newcommand{\rs}{{\cal R}}
\newcommand{\bzk}{BzK}
\newcommand{\kab}{K_{\rm AB}}
\newcommand{\ks}{K_{\rm s}}
\newcommand{\lya}{Lyman~$\alpha$}
\newcommand{\lyb}{Lyman~$\beta$}
\newcommand{\za}{$z_{\rm abs}$}
\newcommand{\ze}{$z_{\rm em}$}
\newcommand{\cmtwo}{cm$^{-2}$}
\newcommand{\nhi}{$N$(H$^0$)}
\newcommand{\degpoint}{\mbox{$^\circ\mskip-7.0mu.\,$}}
\newcommand{\kms}{\,km~s$^{-1}$}      
\newcommand{\minpoint}{\mbox{$'\mskip-4.7mu.\mskip0.8mu$}}
\newcommand{\peryr}{\mbox{$\>\rm yr^{-1}$}}
\newcommand{\secpoint}{\mbox{$''\mskip-7.6mu.\,$}}
\newcommand{\sqdeg}{\mbox{${\rm deg}^2$}}
\newcommand{\squig}{\sim\!\!}
\newcommand{\subsun}{\mbox{$_{\twelvesy\odot}$}}
\newcommand{\et}{{\rm et al.}~}

\def\ltsima{$\; \buildrel < \over \sim \;$}
\def\simlt{\lower.5ex\hbox{\ltsima}}
\def\gtsima{$\; \buildrel > \over \sim \;$}
\def\simgt{\lower.5ex\hbox{\gtsima}}
\def\arcs{$''~$}
\def\arcm{$'~$}
\def\erf{\mathop{\rm erf}}
\def\erfc{\mathop{\rm erfc}}
\title{
A CENSUS OF OPTICAL AND NEAR-INFRARED SELECTED STAR-FORMING
AND PASSIVELY EVOLVING GALAXIES AT REDSHIFT $Z\sim 2$\altaffilmark{1}}
\author{\sc Naveen A. Reddy, Dawn K. Erb, Charles C. Steidel}
\affil{California Institute of Technology, MS 105--24,
Pasadena, CA 91125} 
\author{\sc Alice E. Shapley\altaffilmark{2}}
\affil{Astronomy Department, University of California, Berkeley, 
601 Campbell Hall, Berkeley, CA 94720}
\author{\sc Kurt L. Adelberger\altaffilmark{3}}
\affil{Observatories of the Carnegie Institution of Washington, 
813 Santa Barbara Street, Pasadena, CA 91101}
\and
\author{\sc Max Pettini}
\affil{Institute of Astronomy, Madingley Road, Cambridge CB3 OHA, UK}

\altaffiltext{1}{Based, in part, on data obtained at the W.M. Keck
Observatory, which is operated as a scientific partnership among the
California Institute of Technology, the University of California, and
NASA, and was made possible by the generous financial support of the
W.M. Keck Foundation.}
\altaffiltext{2}{Miller Fellow}
\altaffiltext{3}{Carnegie Fellow}


\begin{abstract}

Using the extensive multi-wavelength data in the GOODS-North field,
including our ground-based rest-frame UV spectroscopy and near-IR
imaging, we construct and draw comparisons between samples of optical
and near-IR selected star-forming and passively evolving galaxies at
redshifts $1.4\la z\la 2.6$.  We find overlap at the $70-80\%$ level
in samples of $z\sim 2$ star-forming galaxies selected by their
optical ($U_{\rm n}G{\cal R}$) and near-IR ($\bzk$) colors when
subjected to common $K$-band limits.  Deep {\it Chandra} data indicate
a $\sim 25\%$ AGN fraction among near-IR selected objects, much of
which occurs among near-IR bright objects ($\ks<20$; Vega). Using
X-rays as a proxy for bolometric star formation rate (SFR) and
stacking the X-ray emission for the remaining (non-AGN) galaxies, we
find the SFR distributions of $U_{\rm n}G{\cal R}$, $\bzk$, and
$\jmk>2.3$ galaxies (i.e., Distant Red Galaxies; DRGs) are very
similar as a function of $\ks$, with $\ks<20$ galaxies having $\langle
SFR\rangle\sim 120$~$\sfr$, a factor of two to three higher than those
with $\ks>20.5$.  The absence of X-ray emission from the reddest DRGs
and $\bzk$ galaxies with $\zmk\ga 3$ indicates they must have
declining star formation histories to explain their red colors and low
SFRs.  While the $M/L$ ratio of passively-evolving galaxies may be
larger on average, the {\it Spitzer}/IRAC data indicate that their
inferred stellar masses do not exceed the range spanned by optically
selected galaxies, suggesting that the disparity in current SFR may
not indicate a fundamental difference between optical and near-IR
selected massive galaxies ($M^* > 10^{11}$~M$_{\odot}$).  We consider
the contribution of optical, near-IR, and submillimeter-selected
galaxies to the star formation rate density (SFRD) at $z\sim 2$,
taking into account sample overlap.  The SFRD in the interval $1.4\la
z\la 2.6$ of $U_{\rm n}G{\cal R}$ and $\bzk$ galaxies to $\ks=22$, and
DRGs to $\ks=21$ is $\sim 0.10\pm 0.02$~$\sfr$~Mpc$^{-3}$.
Optically-selected galaxies to ${\cal R}=25.5$ and $\ks=22.0$ account
for $\sim 70\%$ of this total.  Greater than $80\%$ of radio-selected
submillimeter galaxies to $S_{\rm 850\mu m}\sim 4$~mJy with redshifts
$1.4<z<2.6$ satisfy either one or more of the $\ugr$, $\bzk$, and DRG
criteria.

\end{abstract}


\section{Introduction}

A number of surveys have been developed to select galaxies at $z\sim
2$, determine their bolometric star formation rates (SFRs), and
compare with other multi-wavelength studies to form a census of the
total star formation rate density (SFRD) at $z\sim 2$ (e.g.,
\citealt{steidel04}; \citealt{rubin04}; \citealt{daddi04}).  A
parallel line of study has been to compare optical and near-IR
selected galaxies that are the plausible progenitors of the local
population of passively evolving massive galaxies.  However, biases
inherent in surveys that select galaxies based on their star formation
activity (e.g., \citealt{steidel04}) and stellar mass (e.g.,
\citealt{cimatti02a}; \citealt{glazebrook04}) can complicate such
comparisons.  Only with an accurate knowledge of the overlap between
these samples can we begin to address the associations between
galaxies selected in different ways, their mutual contribution to the
SFRD at $z\sim 2$, and the prevalence and properties of passively
evolving and massive galaxies at high redshift.  Quantifying this
overlap between optical and near-IR surveys is a primary goal of this
paper.

In practice, optical surveys are designed to {\it efficiently} select
galaxies with a specific range of properties.  The imaging required
for optical selection is generally a small fraction of the time
required for near-IR imaging, and can cover much larger areas within
that time.  In contrast, near-IR surveys sample galaxies over a wider
baseline in wavelength than optical surveys, and can include galaxies
relevant to studying both the star formation rate and stellar mass
densities at high redshift.  However, in order to achieve a depth
similar (and area comparable) to that of optical surveys, near-IR
selection requires extremely deep imaging and can be quite expensive
in terms of telescope time due to the relatively small size of IR
arrays compared to CCDs.  Furthermore, the ``color'' of the
terrestrial background for imaging is $(B-\ks)_{\rm AB} \simeq 7$
magnitudes, much redder than all but the most extreme $z\sim 2$
galaxies.  Once selected, of course, such extreme galaxies then
require heroic efforts to obtain spectra, whereas optical selection,
particularly at redshifts where key features fall shortward of the
bright OH emission ``forest'', virtually guarantees that one can
obtain a spectroscopic redshift with a modest investment of 8-10m
telescope time and a spectrograph with reasonably high throughput.  As
we show below, optical and near-IR surveys complement each other in a
way that is necessary for obtaining a reasonably complete census of
galaxies at high redshift.

The SFRs of $z\sim 2$ galaxies are typically estimated by employing
locally-calibrated relations between emission at which the galaxies
can be easily detected (e.g., UV, H$\alpha$) and their FIR emission.
The X-ray luminosity of local non-active galaxies results primarily
from high mass X-ray binaries, supernovae, and diffuse hot gas (e.g.,
\citealt{grimm02}; \citealt{strickland04}); all of these sources of
X-ray emission are related to the star formation activity on
timescales of $\la 100$~Myr.  Observations of galaxies in the local
universe show a tight correlation between X-ray and FIR luminosity,
prompting the use of X-ray emission as an SFR indicator
\citep{ranalli03}.  This correlation between X-ray emission and SFR
applies to galaxies with a very large range in SFRs, from $\sim
0.1-1000$~$\sfr$.  Stacking analyses at X-ray and radio wavelengths,
and comparison with UV emission, indicate that the local SFR relations
appear to give comparable estimates of the instantaneous SFRs of
galaxies after assuming continuous star formation models and
correcting for dust (e.g., \citealt{reddy04}; \citealt{nandra02};
\citealt{seibert02}).

Two surveys designed to select massive galaxies at redshifts $1.4\la
z\la 2.5$ and passively-evolving (PE) galaxies at redshifts $z\ga 2$,
respectively, are the K20 and FIRES surveys.  The K20 and FIRES
selection criteria were developed to take advantage of the sensitivity
of rest-frame optical light and color to stellar mass and the strength
of the Balmer break, respectively, for $z\sim 2$ galaxies (e.g.,
\citealt{cimatti02b}; \citealt{franx03}).  The Gemini Deep Deep Survey
(GDDS) extends this near-IR technique to target massive galaxies at
slightly lower redshifts ($0.8\la z \la 2.0$; \citealt{abraham04}).

X-ray stacking analyses of the brightest galaxies in the K20 and FIRES
surveys indicate an average SFR a factor of $4$ to $5$ times larger
than for optically-selected $z\sim 2$ galaxies (\citealt{daddi04b};
\citealt{rubin04}), inviting the conclusion that optical selection
misses a large fraction of the star formation density at high
redshift.  While it is certainly true that optical surveys miss some
fraction of the SFRD, the past quoted difference in the average SFRs
of galaxies selected optically and in the near-IR disappears once the
galaxies are subjected to a common near-IR magnitude limit, as we show
below.

We have recently concluded a campaign to obtain deep near-IR imaging
for fields in the $z\sim 2$ optical survey \citep{steidel04}, allowing
for a direct comparison of optical and near-IR selected galaxies.  One
result of this comparison is that $\ks<20$ (Vega) optically-selected
galaxies show similar space densities, stellar masses, and
metallicities as $\ks$-bright galaxies in near-IR samples
\citep{shapley04}.  More recently, \citet{adelberger2005} show that
the correlation lengths for $\ks$-bright galaxies among optical and
near-IR samples are similar, suggesting an overlap between the two
sets of galaxies, both of which plausibly host the progenitors of
massive elliptical galaxies in the local universe.  These results
suggest that near-IR bright galaxies have similar properties
regardless of the method used to select them.

In this paper, we extend these results by examining the color
distributions and X-ray properties of near-IR and optically selected
galaxies at $z\sim 2$ in the GOODS-North field \citep{giavalisco04}.
The field is well-suited for this analysis given the wealth of
complementary data available, including {\it Chandra}/X-ray,
ground-based optical and near-IR, and {\it Spitzer}/IRAC imaging.
Multi-wavelength data in a single field are particularly useful in
that we can use a common method for extracting photometry that is not
subject to the biases that may exist when comparing galaxies in
different fields whose fluxes are derived in different ways.  The
addition of our rest-frame UV spectroscopic data in the GOODS-N field
provides for a more detailed analysis than otherwise possible of the
properties of galaxies as a function of selection technique.
Furthermore, the GOODS-N field coincides with the {\it Chandra} Deep
Field North (CDF-N) region which have the deepest ($2$~Ms) X-ray data
available \citep{alexander03}.  The X-ray data allow for an
independent estimate of bolometric SFRs and the available depth allows
more leeway in stacking smaller numbers of sources to obtain a
statistical detection, as well as identifying AGN to a lower
luminosity threshold than possible in other fields that have shallower
X-ray data.

The outline of the paper is as follows.  In \S 2, we describe the
optical, near-IR, X-ray, and IRAC data and present the optical and
near-IR selection criteria and X-ray stacking method.  Color
distributions, direct X-ray detections, and stacked results are
examined in \S 3.  In \S 4, we discuss the SFR distributions of
optical and near-IR selected $z\sim 2$ galaxies and their relative
contributions to the SFRD, and the presence of a passively evolving
population of galaxies.  A flat $\Lambda$CDM cosmology is assumed with
$H_{0}=70$~km~s$^{-1}$~Mpc$^{-1}$ and $\Omega_{\Lambda}=0.7$.

\section{Data and Sample Selection}

\subsection{Imaging}
\label{sec:imaging}

\begin{deluxetable*}{lcccc}[!tbp]
\tabletypesize{\footnotesize}
\tablewidth{0pc}
\tablecaption{Interloper Contamination of the $\ugr$ Sample}
\tablehead{
\colhead{$\ks$ Range} &
\colhead{$N_{\rm phot}$\tablenotemark{a}} &
\colhead{$N_{\rm spec}$\tablenotemark{b}} &
\colhead{$N_{\rm z<1}$\tablenotemark{c}} &
\colhead{$f_{\rm z<1}$\tablenotemark{d}}}
\startdata
$\ks\leq 20.0$ & 61 & 18 & 7 & 0.39\\
$20.0<\ks\leq20.5$ & 58 & 23 & 2 & 0.09 \\
$20.5<\ks\leq21.0$ & 82 & 30 & 3 & 0.10 \\
$21.0<\ks\leq21.5$ & 101 & 32 & 2 & 0.06 \\
$21.5<\ks\leq22.0$ & 141 & 29 & 3 & 0.10 \\
\enddata
\tablenotetext{a}{Number of photometric $\ugr$ candidates.}
\tablenotetext{b}{Number of candidates with spectroscopic redshifts.}
\tablenotetext{c}{Number of interlopers.}
\tablenotetext{d}{Interloper fraction.}
\label{tab:kinterloper}
\end{deluxetable*}

Optical $U_{\rm n}G{\cal R}$ images in the GOODS-North field were
obtained in 2002 and 2003 April under photometric conditions using the
KPNO and Keck I telescopes.  The KPNO/MOSAIC $U$-band image was
obtained from the GOODS team (PI: Giavalisco) and was transformed to
reflect $U_{\rm n}$ magnitudes (e.g., \citealt{steidel04}).  The Keck
I $G$ and ${\cal R}$ band images were taken by us with the Low
Resolution Imaging Spectrograph (LRIS; \citealt{oke95},
\citealt{steidel04}), and were oriented to provide the maximum overlap
with the GOODS ACS and {\it Spitzer} survey region.  The images cover
$11\arcmin~\times~15\arcmin$ with FWHM $\sim 0\farcs 7$ to a depth of
${\cal R}\sim 27.5$ ($3$~$\sigma$).  Image reduction and photometry
were done following the procedures described in \citet{steidel03}.  We
obtained deep $B$-band images of the GOODS-N field from a public
distribution of $Subaru$ data \citep{capak04}.  The deep $z$-band data
are acquired from the public distribution of the $HST$ Advanced Camera
for Surveys (ACS) data \citep{giavalisco04}.  The $B$ and $z$ band
data have $5~\sigma$ depth of $26.9$ and $27.4$~mag measured in
$3$~$\arcsec$ and $0\farcs2$ diameter apertures, respectively.  The
$\ks$ and $J$ imaging was accomplished with the Wide Field Infrared
Camera (WIRC) on the Palomar Hale $5$~m telescope \citep{wilson03},
providing $8\farcm7~\times~8\farcm7$ coverage in the central portion
of the GOODS-N field.  The near-IR images cover $\sim 43\%$ of the
optical image.  The images had FWHM $\sim 1\farcs 0$ under photometric
conditions and $3$~$\sigma$ sensitivity limits of $\sim 22.6$ and
$\sim 24.1$~mag in the $\ks$ and $J$ bands, respectively.  The near-IR
data are described in detail by \citet{erb05}.  The total area studied
in the subsequent analysis is $\sim 72.3$~arcmin$^{2}$.

The procedures for source detection and photometry are described in
\citet{steidel03}.  Briefly, $U_{\rm n}G{\cal R}$ magnitudes were
calculated assuming isophotal apertures that were adjusted to the
${\cal R}$-band flux profiles.  Source detection was done at
$\ks$-band.  $\bzk$ and $J$ magnitudes are computed assuming the
isophotal apertures adjusted to the $\ks$-band flux profiles, unless
the ${\cal R}$-band isophotes gave a more significant $\ks$ detection.
In the analysis to follow, ``$\ks$'' and $J$ magnitudes are in Vega
units.  We use the conversion $K_{\rm AB} = \ks + 1.82$.  All other
magnitudes are in AB units.

Fully reduced {\it Spitzer}/IRAC mosaics of the GOODS-North field were
made public in the first data release of the GOODS Legacy project (PI:
Dickinson).  The IRAC data overlap completely with our $\ks$-band
image, but currently only two channels (either $3.6$~$\mu$m and
$5.8$~$\mu$m, or $4.5$~$\mu$m and $8.0$~$\mu$m) are available over
most of the image.  A small area of overlap has coverage in all four
channels.  The images are deep enough that source confusion is an
issue.  We have mitigated the effects of confusion noise by employing
the higher spatial resolution $\ks$-band data to constrain source
positions and deblend confused IRAC sources.  We performed PSF
photometry using the procedure described in \citet{shapley05}.

\subsection{Selection Criteria}

\subsubsection{Optical Selection of Star-Forming Galaxies}

We have optically-selected $z\sim 2$ galaxies in the GOODS-N field
based on their observed $U_n G {\cal R}$ colors (\citealt{adel04};
\citealt{steidel04}) to a limiting magnitude of ${\cal R}=25.5$.  The
selection criteria aim to select actively star forming galaxies at
$z\sim 2$ with the same range in UV properties and extinction as LBGs
at $z\sim 3$ \citep{steidel03}.  ``BX'' galaxies are selected to be at
redshifts $2.0\la z\la 2.6$ using the following criteria:
\begin{eqnarray}
G-{\cal R} &\geq& -0.2\nonumber\\
U_n-G      &\geq& G-{\cal R}+0.2\nonumber\\
G-{\cal R} &\leq &0.2(U_n-G)+0.4\nonumber\\
U_n-G      &\leq& G-{\cal R}+1.0,
\end{eqnarray}
and ``BM'' objects are selected to be at redshifts $1.5\la z \la 2.0$ using
the following criteria:
\begin{eqnarray}
G-{\cal R} &\geq& -0.2\nonumber\\
U_n-G      &\geq& G-{\cal R}-0.1\nonumber\\
G-{\cal R} &\leq &0.2(U_n-G)+0.4\nonumber\\
U_n-G      &\leq& G-{\cal R}+0.2
\end{eqnarray}
(\citealt{adel04}; \citealt{steidel04}).  
For subsequent analysis, we will refer to BX and BM objects as those
which are optically-, or ``$\ugr$''-, selected.

\begin{figure*}[!tbh]
\plotone{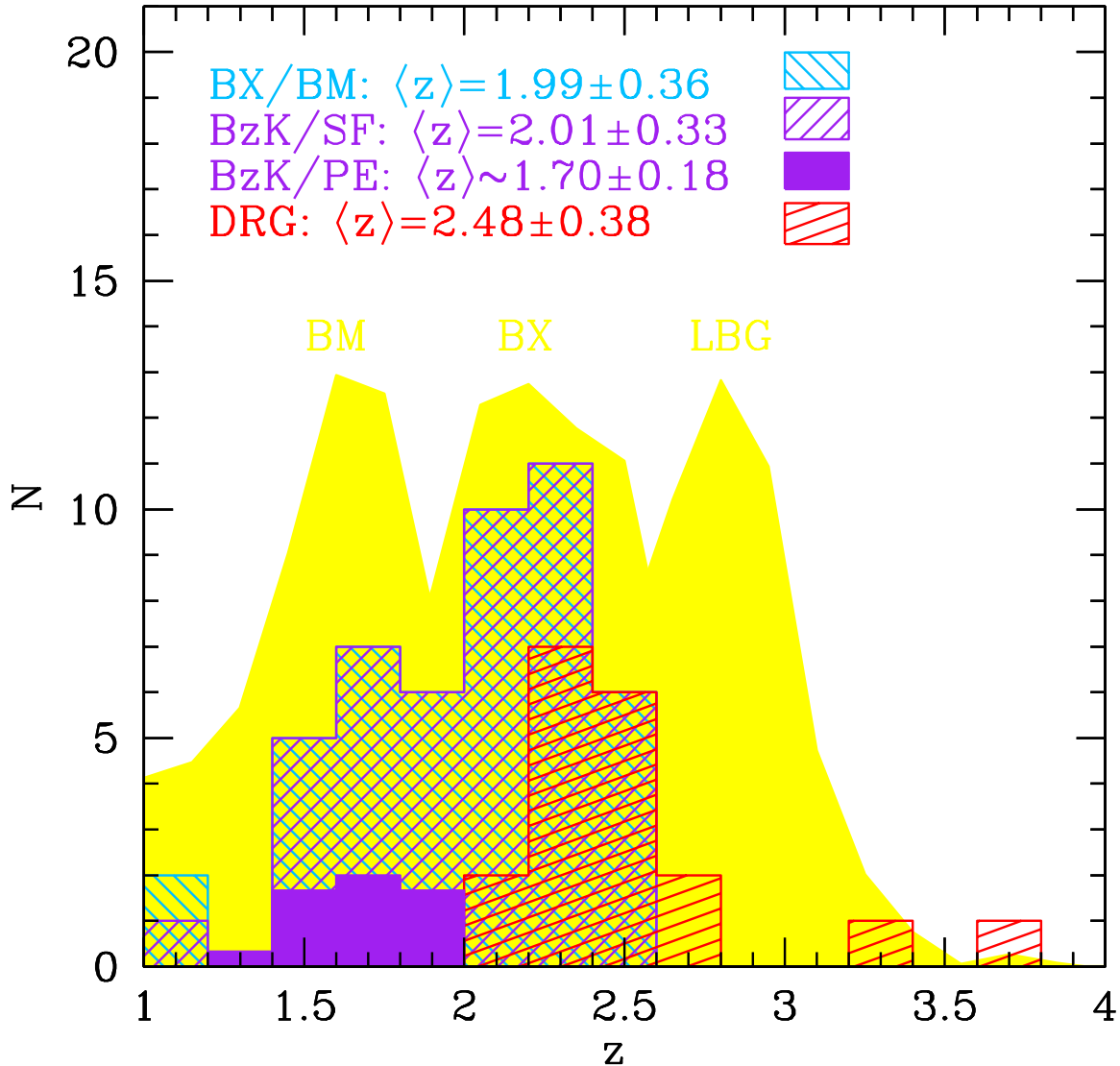}
\caption{{\it Spectroscopic} redshift distributions to $\ks=21$ for
  the various samples considered here.  The $\ugr$ and $\bzk$/SF
  distributions (hashed histograms) include sources from our sample in
  the GOODS-North field and overlap almost completely.  The DRGs have
  a higher mean redshift of $\langle z\rangle=2.48\pm0.38$ from our
  sample of $\jmk>2.3$ sources with $z>1$ in all four fields of the
  optical survey \citep{steidel04} where we have complementary $J$ and
  $K$-band imaging.  The redshift distribution of DRGs within our
  sample (all of which are selected with the $\ugr$ or $z\sim 3$ LBG
  criteria) is similar to that found by \citet{vandokkum04},
  \citet{vandokkum03}, and \citet{forster04}.  The solid histogram
  shows the redshift distribution for $\bzk$/PE galaxies from
  \citet{daddi04b} and \citet{daddi05}, scaled down by a factor of $3$
  for clarity.  The background shaded regions show the arbitrarily
  normalized redshift distributions for optically-selected BX and BM
  galaxies, and LBGs.
\label{fig:zhist}}
\end{figure*}

Optical color selection of $z\sim 2$ galaxies in the $11\arcmin$ by
$15\arcmin$ area of the GOODS-North field yielded 1360 BX and BM
candidates, of which 620 lie in the region where we have complementary
$J$ and $K$-band data (\S~\ref{sec:imaging}), and $199$ have
$\ks<21.0$.  Followup spectroscopy with the blue channel of the Low
Resolution Imaging Spectrograph (LRIS-B) yielded $147$ redshifts for
objects with $\ks$-band data ($248$ redshifts over the entire optical
field).  Of these $147$ objects with redshifts and $\ks$-band data,
$129$ have $z>1$, and $60$ have $z>1$ {\it and} $\ks<21$.  The mean
redshift of the $60$ $\ugr$ objects is $\langle z\rangle=1.99\pm0.36$.
The spectroscopic interloper fractions in the $\ugr$ sample are
summarized in Table~\ref{tab:kinterloper}.  The BX and BM selection
functions (shown as shaded distributions in Figure~\ref{fig:zhist})
have distributions with mean redshifts $\langle z\rangle =2.2\pm 0.3$
and $\langle z\rangle =1.7\pm 0.3$, respectively \citep{steidel04},
and the combination of these two samples comprise our $\ugr$-selected
$z\sim 2$ sample.  In the analysis to follow, we designate an
interloper as any object with $z<1$.

\subsubsection{Near-IR Selection of Star-Forming Galaxies}

\begin{figure}[hbt]
\centerline{\epsfxsize=8.5cm\epsffile{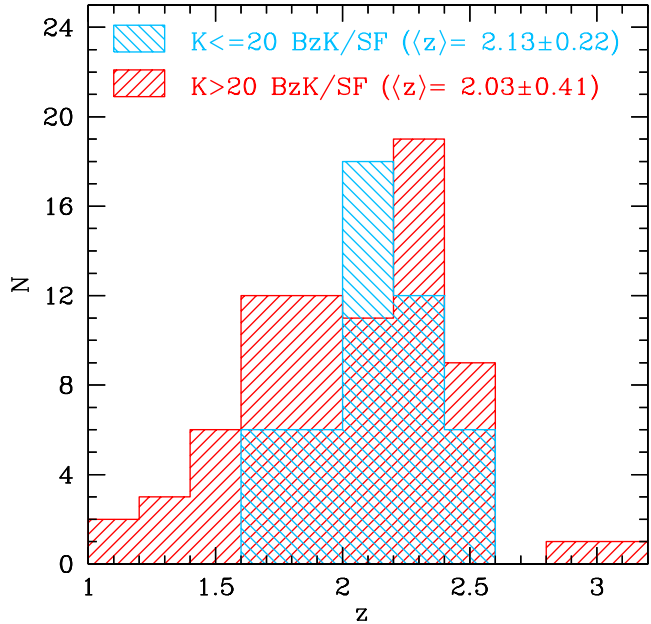}}
\figcaption[f2.eps]{Arbitrarily normalized spectroscopic redshift distribution of
  $\bzk$/SF galaxies in our spectroscopic sample to $\ks\sim 22.5$,
  with separate emphasis on $\ks\le 20$ and $\ks>20$ $\bzk$/SF
  objects.  The $\bzk$/SF criteria select $\ks>20$ objects over a
  broader range in redshift than $\ks\le 20$ objects.
\label{fig:nzbzk}}
\end{figure}

The near-IR properties of galaxies can be used both to target star
forming galaxies and to identify those with extremely red colors which
may indicate passive evolution.  To address the former issue, we have
employed the ``$\bzk$'' selection criteria of \citet{daddi04b} to
cull objects in the GOODS-N field and directly compare with those
selected on the observed optical properties of $z\ga 2$ galaxies.
\citet{daddi04b} define the quantity ``$\bzk$'':
\begin{eqnarray}
BzK \equiv (z-K)-(B-z);
\end{eqnarray}
star-forming galaxies with $z>1.4$ are targeted
by the following criterion:
\begin{eqnarray}
BzK \ge -0.2,
\label{eq:bzkeq}
\end{eqnarray}
in AB magnitudes.  Of the $1185$ sources with $>3$~$\sigma$ $B$, $z$,
and $K$ detections and $\ks<21$, 221 satisfy Equation~\ref{eq:bzkeq}.
The surface density of $\bzk$ galaxies with $\ks<21$ is $\sim
3$~arcmin$^{-2}$, similar to the surface density of $\ugr$ galaxies to
a similar $\ks$-band depth.  These star-forming $\bzk$ galaxies will
be referred to as ``$\bzk$/SF'' galaxies, and their spectroscopic
redshift distribution {\it from our spectroscopic sample} is shown in
Figure~\ref{fig:zhist}.  Our deep near-IR imaging allows us to
determine the redshift distribution for $\bzk$/SF galaxies with
$\ks>20$ (and which also satisfy the $\ugr$ criteria), and the results
are shown in Figure~\ref{fig:nzbzk}.  The mean redshifts of the
$\ks\le 20$ and $\ks>20$ distributions are $\langle z\rangle = 2.13\pm
0.22$ and $\langle z\rangle = 2.03\pm 0.41$, respectively, and agree
within the uncertainty.  We note, however, that the $\bzk$/SF criteria
select $\ks>20$ objects over a broader range in redshift ($1.0\le z
\le 3.2$) than $\ks\le 20$ objects.  This reflects the larger range in
$\bzk$ colors of $\ks>20$ $\bzk$/SF galaxies compared with those
having $\ks\le 20$.  Additionally, the photometric scatter in colors
is expected to increase for fainter objects, so a broadening of the
redshift distribution for $\bzk$/SF objects with fainter $\ks$
magnitudes is not surprising.

We emphasize that we only know the redshifts for $\bzk$/SF galaxies
which also happen to fall in the $\ugr$ sample.  In general, the true
redshift distribution, $N_{o}^{\bzk/SF}(z)$, of the $\bzk$/SF sample
will be broader than the distributions shown in Figure~\ref{fig:zhist}
and Figure~\ref{fig:nzbzk}, call them $N_{c}^{\bzk/SF}(z)$), which are
effectively convolved with the $\ugr$ selection function.  For
example, the rapid dropoff in $N_{c}^{\bzk/SF}(z)$ for $z>2.6$
(Figure~\ref{fig:nzbzk}) may simply reflect the dropoff in the BX
selection function for $z>2.6$.  However, the $N_{c}^{\bzk/SF}(z)$ we
derive here is similar to that of the {\it photometric} redshift
distribution of K20 galaxies from \citet{daddi04b}, which is subject
to its own systematic errors, suggesting that a reasonable
approximation is to take $N_{o}^{\bzk/SF}(z) \simeq
N_{c}^{\bzk/SF}(z)$.

\subsubsection{Near-IR Selection of Passively Evolving Galaxies}

\begin{figure}[hbt]
\centerline{\epsfxsize=8.5cm\epsffile{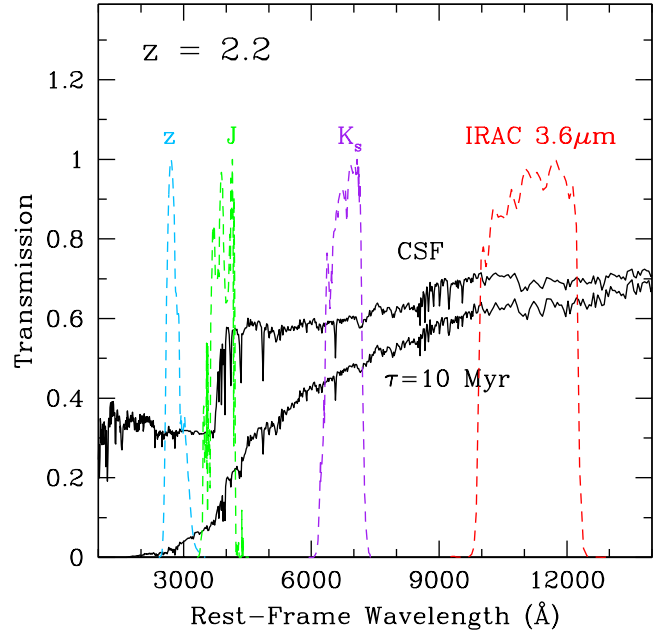}}
\figcaption[f3.eps]{Relative transmission of the $z$, $J$, $\ks$, and IRAC
  $3.6$~$\mu$m filters at rest-frame wavelengths for $z=2.2$.  Also
  shown are typical (unreddened) galaxy SEDs assuming constant star
  formation (CSF) and instantaneous star formation ($\tau=10$~Myr)
  aged to $1$~Gyr.  For redshifts $z\sim 1.88-2.38$, the $J$-band
  brackets the prominent Balmer and $4000$~\AA\, breaks.
\label{fig:sed}}
\end{figure}

\begin{figure}[thb]
\centerline{\epsfxsize=8.5cm\epsffile{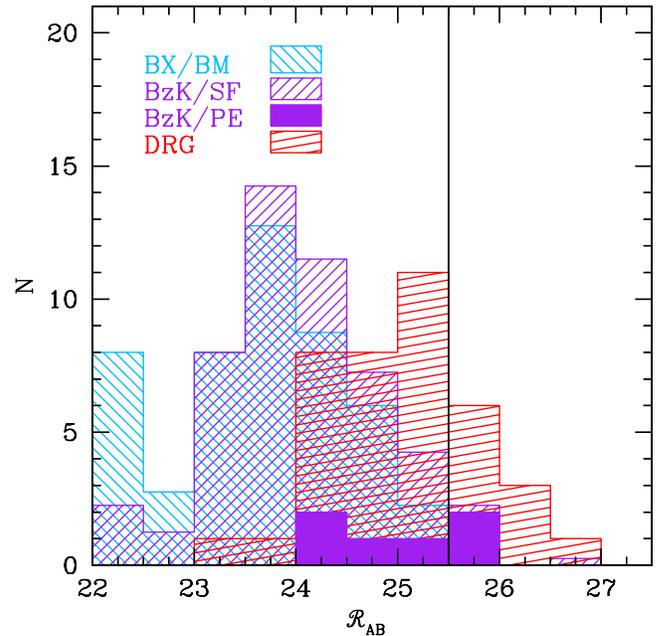}}
\figcaption[f4.eps]{Optical magnitude distributions for photometrically selected
  $\ks<21$ galaxies in the $\ugr$, $\bzk$, and DRG samples.  The solid
  vertical line denotes the magnitude limit for galaxies in the
  optically-selected ($\ugr$) sample.  Approximately $47\%$ of DRGs
  ($34/73$) have ${\cal R}>27.0$ and are not shown in the figure.  The
  distribution of $\bzk$/PE galaxies has been arbitrarily normalized
  for clarity.
\label{fig:rdist}}
\end{figure}

\begin{deluxetable*}{llcccccc}[bht]
\tabletypesize{\footnotesize}
\tablewidth{0pc}
\tablecaption{Sample Properties}
\tablehead{
\colhead{} &
\colhead{} &
\colhead{} &
\colhead{} &
\colhead{} &
\colhead{} &
\colhead{$\langle L_{\rm 2.0-10\, keV}\rangle$} &
\colhead{$\langle \rm SFR_{\rm x}\rangle$}  \\
\colhead{$\ks$ Range} &
\colhead{Sample} &
\colhead{$N_{\rm T}$\tablenotemark{a}} &
\colhead{$N_{\rm X}$\tablenotemark{b}} & 
\colhead{$N_{\rm S}$\tablenotemark{c}} &
\colhead{$\langle z\rangle$\tablenotemark{d}} &
\colhead{($\times 10^{41}$~ergs~s$^{-1}$)} &
\colhead{(M$_{\odot}$~yr$^{-1}$)}}
\startdata
$18.0<\ks\leq 20.0$ & $\ugr$\tablenotemark{e} & 11 & 4 & 10 (7) & 1.75 (1.80) & $7.13\pm0.88$ ($4.95\pm1.15$) & 143 (99) \\
                    & $\bzk$/SF & 77 & 24& 45 (42)& 1.97 (2.01) & $5.39\pm0.46$ ($4.68\pm0.48$) & 108 (94) \\
                    & $\bzk$/PE & 14 & 3 & 11     & 1.74        & $3.07\pm0.84$ & 61 \\
                    & DRG       & 20 & 5 & 14 & 2.48 & $5.26\pm1.28$ & 105 \\
\\
$20.0<\ks\leq 20.5$ & $\ugr$ \tablenotemark{e}   & 21 & 0 & 21     & 2.03        & $4.89\pm0.79$                 & 98       \\
                    & $\bzk$/SF & 57 & 5 & 56 (55)& 2.01 (2.01) & $3.96\pm0.40$ ($3.87\pm0.41$) & 79 (77)  \\
                    & $\bzk$/PE & 0  & 0 & ... & ... & ... & ...\\
                    & DRG       & 20 & 7 & 13 & 2.48 & $2.32\pm1.30$ & 46 \\ 
\\
$20.5<\ks\leq 21.0$ & $\ugr$ \tablenotemark{e}   & 27 & 0 & 27     & 1.99        & $2.11\pm0.56$                 & 42       \\
                    & $\bzk$/SF & 87 & 3 & 82 (81)& 2.01 (2.01) & $2.32\pm0.30$ ($2.20\pm0.30$) & 46 (44) \\
                    & $\bzk$/PE & 3  & 1 & 2 & 1.74 & $2.07\pm1.54$ & 41 \\
                    & DRG       & 33 & 7 & 26& 2.48 & $1.90\pm0.86$ & 38 \\
\\
$21.0<\ks\leq 21.5$ & $\ugr$ \tablenotemark{e} & 31 & 0 & 31 & 2.04 & $2.95\pm1.01$ & 59 \\
                    & $\bzk$/SF & 99 & 2 & 97 & 2.01 & $2.76\pm 0.28$ & 55 \\
\\
$21.5<\ks\leq 22.0$ & $\ugr$ \tablenotemark{e} & 26 &  0 & 26 & 2.22 & $1.43\pm0.77$ & 29 \\
                    & $\bzk$/SF & 148 & 0 & 148 & 2.01 & $2.78\pm 0.73$ & 56 \\
\\
$22.0<\ks\leq 22.5$ & $\ugr$ \tablenotemark{f} & 93 & 0 & 93 & 2.04 & $1.13\pm0.25$ & 23 \\
                    & $\bzk$/SF & 77 & 0 & 77 & 2.01 & $0.49\pm0.29$ & 10 \\
\enddata
\tablenotetext{*}{Note.---Values in parentheses are when we exclude all directly-detected
X-ray sources, including ones which may be star-forming galaxies (Table~\ref{tab:liksf}).}
\tablenotetext{a}{Total number of sources in sample.}
\tablenotetext{b}{Number of direct X-ray detections, corresponding to a minimum $3$~$\sigma$ flux
of $f_{\rm 0.5-2.0~keV} \sim 2.5\times 10^{-17}$~erg~s$^{-1}$~cm$^{-2}$.}
\tablenotetext{c}{Number of stacked sources.}
\tablenotetext{d}{Mean redshift of stacked sample.  Sources without a spectroscopic redshift
are assigned the mean redshift of the sample to which they belong, where the mean redshifts for
the sample are specified in Figure~\ref{fig:zhist}.}
\tablenotetext{e}{We only consider $\ugr$ galaxies which are spectroscopically confirmed to 
lie at redshifts $z>1$, $\sim 25\%$ of the photometric sample of $\ugr$ galaxies.}
\tablenotetext{f}{Photometric $\ugr$ galaxies.}
\label{tab:ldist}
\end{deluxetable*}

In addition to the criteria above, several methods have been developed
to select passively evolving high redshift galaxies by exploiting the
presence of absorption or continuum breaks in the SEDs of galaxies
with dominant old stellar populations.  The $\bzk$ selection criteria
\begin{eqnarray}
BzK & < & -0.2\nonumber\\
z-K & > & 2.5
\end{eqnarray}
are designed to select passively evolving galaxies at $z>1.4$
(\citealt{cimatti04}; \citealt{daddi04b}).  One galaxy which has a
secure $B$-band detection, and an additional $16$ with $B$-band
limits, satisfy these criteria, implying a surface density of
$\bzk$/PE galaxies of $0.24$~arcmin$^{-2}$ to $\ks=21$.  Galaxies
selected by their $\bzk$ colors to be passively-evolving are referred
to as ``$\bzk$/PE'' objects.  The redshift distribution of $\bzk$/PE
galaxies, taken from the spectroscopic samples of \citet{daddi04b} and
\citet{daddi05}, shows that they mostly lie between redshifts $1.4\la
z\la 2$ (Figure~\ref{fig:zhist}).  We note that we may be incomplete
for the $\bzk$/PE objects despite the very deep $B$-band data
considered here and these missing objects may be more easily selected
using the $\jmk>2.3$ criteria discussed below (see also \S~4.2.3).

The $\jmk$ color probes the age-sensitive Balmer and $4000$~\AA\,
breaks for galaxies with redshifts $2.0\la z\la4.5$
(Figure~\ref{fig:sed}).  The criterion
\begin{eqnarray}
J-K_{\rm s} > 2.3
\end{eqnarray}
\citep{franx03} can be used to select both passively evolving and
heavily reddened star-forming galaxies with $\ebmv>0.3$.  Galaxies
satisfying this criterion are also referred to as Distant Red Galaxies
(DRGs; \citealt{franx03}; \citealt{vandokkum04}).  There are $62$
galaxies with $\jmk>2.3$ that are detected in $J$, and an additional
$11$ have $J$-band limits.  The observed surface density of DRGs is
$1.01\pm0.12$~arcmin$^{-2}$ to $\ks=21$, in very good agreement with
the surface density found by \citet{vandokkum04} and
\citet{forster04}.  The {\it spectroscopic} redshift distribution of
DRGs from the four fields of the optical survey where we have deep $J$
and $\ks$-band imaging is shown in Figure~\ref{fig:zhist}, and is
consistent with the redshift distributions found by
\citet{vandokkum04}, \citet{vandokkum03}, and \citet{forster04}.
Star-forming and passively-evolving DRGs are referred to as ``DRG/SF''
and ``DRG/PE'', respectively.  The depth of our $J$-band image implies
that our sample of DRGs will be incomplete for those with $\ks>21$.
Therefore, we have limited ourselves to galaxies with $\ks<21$ when
comparing DRGs with $\bzk$ and/or $\ugr$ selected galaxies.  We
reconsider $\ugr$ and $\bzk$ galaxies with $\ks>21$ as noted below.

The optical magnitude distributions for galaxies with $\ks<21$ are
shown in Figure~\ref{fig:rdist}.  The catalog of $\ugr$ galaxies is
restricted to $\rs<25.5$.  However, our optical imaging is
significantly deeper ($\rs=27.5$; $3$~$\sigma$), allowing us to
extract optical magnitudes for galaxies much fainter than those in the
$\ugr$ catalog.  Most of those galaxies with ${\cal R}>25.5$ are DRGs.
The nature of optically-faint DRGs is discussed in \S~\ref{sec:pass}.

For most of the analysis that follows, we either use only the
spectroscopically confirmed sub-sample of $\ugr$ galaxies, or we apply
our knowledge of the contamination fraction of the photometric sample
to deduce any inferred quantities.  The small available spectroscopic
samples using the near-IR criteria prevent us from applying similar
corrections when deducing properties for the near-IR samples.

\subsection{X-ray Data and Stacking Method}

One focus of this paper is to draw comparisons between galaxies
selected by the techniques described above by using their stacked
X-ray emission as a proxy for their bolometric SFRs.  X-ray stacking
allows us to determine instantaneous bolometric SFRs in a manner that
is independent of extinction and the degeneracies associated with
stellar population modeling.  For example, the average reddening of
rest-frame UV selected galaxies of $\ebmv\sim 0.15$ implies a column
density of $N_{\rm H}\sim 7.5\times 10^{20}$~cm$^{-2}$, assuming the
Galactic calibration \citep{diplas94}.  Absorption in the rest-frame
$2-10$~keV band is negligible for these column densities.  The X-ray
data are taken from the $Chandra$ 2~Ms survey of the GOODS-N field
\citep{alexander03}.  We made use primarily of the soft band (SB;
$0.5-2.0$~keV) data for our analysis, but we also include hard band
(HB; $2.0-8.0$~keV) data to examine the nature of directly-detected
X-ray sources.  The data are corrected for vignetting, exposure time,
and instrumental sensitivity in producing the final mosaicked image.
The final product has an SB on-axis sensitivity of $\sim 2.5\times
10^{-17}$~erg~s$^{-1}$~cm$^{-2}$ ($3$~$\sigma$), sufficient to
directly detect $L_{\rm 2-10~keV} > 9.3\times 10^{41}$~ergs~s$^{-1}$
objects at $z\sim 2$, corresponding to an SFR of $\sim
190$~M$_\odot$~yr$^{-1}$.

The stacking procedure followed here is the same as that discussed in
\citet{reddy04}.  Apertures used to extract X-ray fluxes had radii set
to $2\farcs 5$ for sources within $6\arcmin$ of the average {\it
Chandra} pointing origin, and set to the $50\%$ encircled energy
radius for sources with off-axis angles greater than $6\arcmin$
\citep{feigelson02}.  X-ray fluxes were computed by adding the counts
within apertures randomly dithered by $0\farcs 5$ at the galaxy
positions.  Background estimates were computed by randomly placing the
same sized apertures within $5\arcsec$ of the galaxy positions,
careful to prohibit the placing of a background aperture on a known
X-ray detection.  This procedure of placing random apertures was
repeated 1000 times.  The mean X-ray flux of a galaxy is taken to be
the average of all the flux measurements from the $0\farcs 5$ dithered
apertures and the background noise is taken to be the dispersion in
fluxes measured from the background apertures.  We applied aperture
corrections to the fluxes and assumed count rate to flux conversions
based on the results compiled in Table~7 of \citet{alexander03}, a
photon index $\Gamma = 2.0$, and a Galactic absorption column density
of $N_{\rm H} = 1.6\times 10^{20}$~cm$^{-2}$ \citep{stark92}.  Poisson
errors dominate the uncertainties in flux.

\section{Results}

\subsection{Direct X-ray Detections}
\label{sec:xray}

\begin{figure}[hbt]
\centerline{\epsfxsize=8.5cm\epsffile{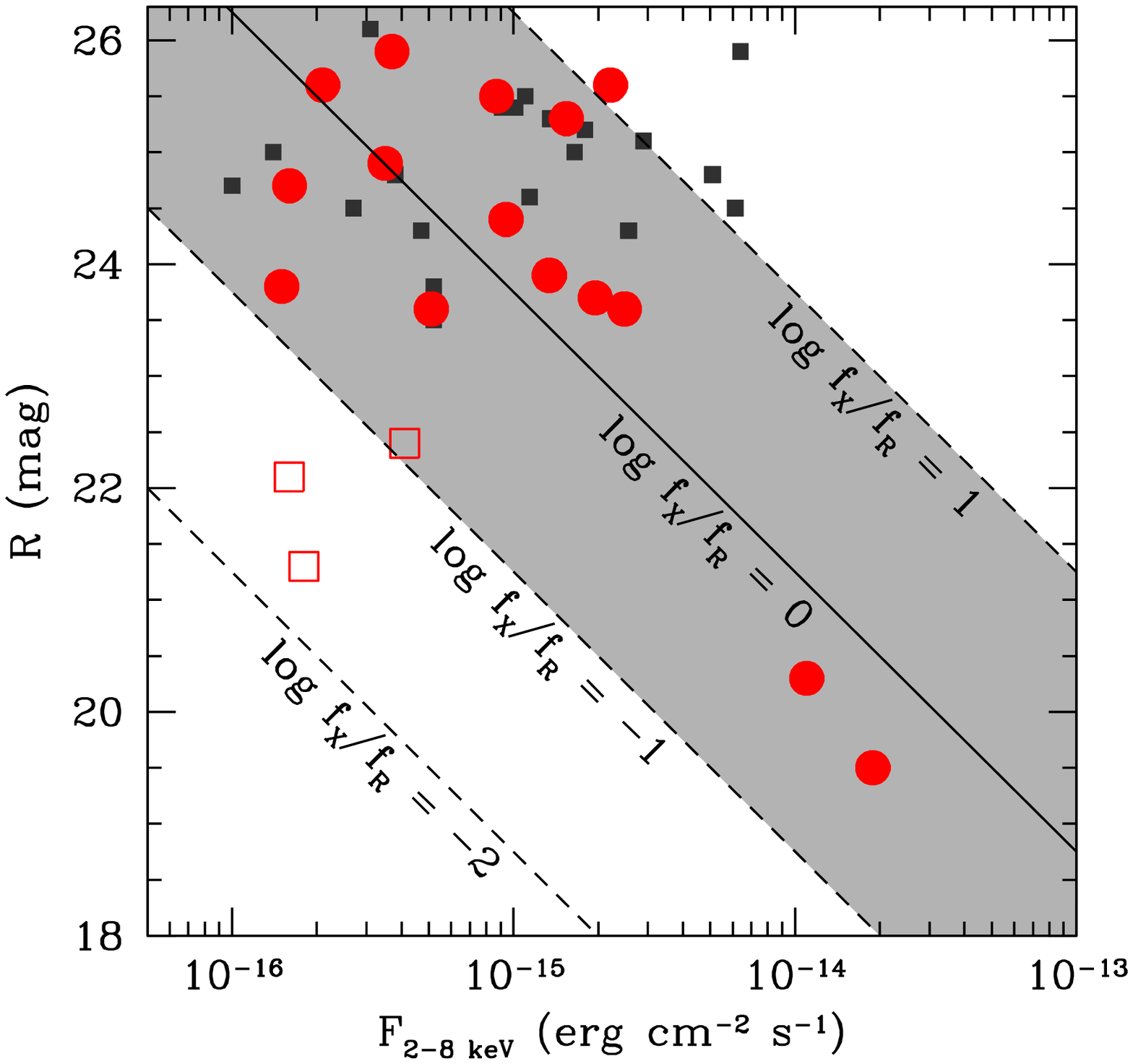}}
\figcaption[f5.eps]{Optical/X-ray flux ratios, defined as $\log f_{\rm X}/f_{\rm
  R} = \log f_{\rm X} +5.50 + R/2.5$ \citep{hornschemeier01}, for all
  directly detected hard band X-ray sources in the $\ugr$, $\bzk$
  (SF/PE), and DRG samples.  The abscissa is the observed hard-band
  flux, corresponding to rest-frame energies of $6-24$~keV, and the
  ordinate is the observed {\it Cousins} R magnitude from the
  compilation of \citet{barger03}.  Large circles denote sources with
  spectroscopic redshifts $z>1$, and are likely AGNs given their
  direct hard band detections.  Sources with hard band detections but
  no redshift identification are shown by the small squares.  Almost
  all sources have optical/X-ray flux ratios between $-1<\log f_{\rm
  X}/f_{\rm R}<1$ (shaded region), values commonly found for AGNs.
  Those with $\log f_{\rm X}/f_{\rm R}<-1$ are confirmed interlopers,
  shown by the large open squares.  Starburst galaxies generally have $\log
  f_{\rm X}/f_{\rm R}<-1$.
\label{fig:xo}}
\end{figure}

\begin{deluxetable*}{lccccccc}[thb]
\tabletypesize{\footnotesize}
\tablewidth{0pc}
\tablecaption{Possible Star-Forming Direct X-ray Detections}
\tablehead{
\colhead{$\alpha$} &
\colhead{$\delta$} &
\colhead{} &
\colhead{} &
\colhead{$f_{\rm 0.5-2.0~keV}$\tablenotemark{a}} &
\colhead{$\ks$} &
\colhead{R\tablenotemark{a}} &
\colhead{} \\
\colhead{(J2000.0)} &
\colhead{(J2000.0)} &
\colhead{Sample} &
\colhead{$z$} &
\colhead{($\times 10^{-15}$~erg~s$^{-1}$~cm$^{-2}$)} &
\colhead{(Vega mag)} &
\colhead{(mag)} &
\colhead{$\log f_{\rm X}/f_{\rm R}$\tablenotemark{b}}}
\startdata
12:36:21.95 & 62:14:15.5 & $\bzk$/SF & 1.38 & 0.02 & 18.95 & 23.1 & -1.96 \\
12:36:52.75 & 62:13:54.8 & $\ugr$ & 1.36 & 0.03 & 19.54 & 22.1 & -2.18 \\
12:36:53.46 & 62:11:40.0 & $\ugr$;$\bzk$/SF & ...  & 0.11 & 18.65 & 22.7 & -1.38 \\
12:36:56.89 & 62:11:12.1 & $\bzk$/SF & ... & 0.02 & 20.50 & 23.8 & -1.68 \\
12:37:03.70 & 62:11:22.6 & $\ugr$;$\bzk$/SF & 1.72 & 0.04 & 19.92 & 23.4 & -1.54 \\
\enddata
\tablenotetext{a}{Soft-band fluxes are from \citet{alexander03} and {\it Cousins} R magnitudes are from
\citet{barger03}.}
\tablenotetext{b}{Defined as $\log f_{\rm X}/f_{\rm
  R} = \log f_{\rm X} +5.50 + R/2.5$ \citep{hornschemeier01}.}
\label{tab:liksf}
\end{deluxetable*}

Of the 221 $\bzk$/SF candidates with $\ks<21$, 32 ($14\%$) have an
X-ray counterpart within $1\farcs 5$ \citep{alexander03}, with a $\sim
0.22\%$ probability for chance superposition.  The X-ray detection
fractions are $24\%$, $6\%$, and $26\%$, for the $\bzk$/PE, $\ugr$,
and DRG samples, respectively, and are summarized in
Table~\ref{tab:ldist}.  Eleven of the $36$ directly-detected
$\bzk$ sources ($32$ in the $\bzk$/SF sample and $4$ in the $\bzk$/PE sample) 
would not have been detected with the sensitivity
of the shallower 1~Msec data in the GOODS-South field studied by
\citet{daddi04b}.  After taking into account the sensitivity
difference, we find a direct X-ray detection rate comparable to that
of \citet{daddi04b} of $\sim 11\%$.  Figure~\ref{fig:xo} shows the
X-ray/optical flux ratios ($\log f_{\rm X}/f_{\rm R}$) for sources in
all four samples ($\ugr$, $\bzk$/SF, $\bzk$/PE, and DRG)
directly-detected in the {\it Chandra} hard band ($2-8$~keV).  Direct
hard band detections must be AGN if they are at $z\sim 2$, since
starburst galaxies with no accretion activity are expected to have
little flux at rest-frame energies of $6-20$~keV.  Indeed, the
X-ray/optical flux ratios for directly detected hard band sources lie
in the region typically populated by AGNs (shaded area of
Figure~\ref{fig:xo}).  A smaller fraction of galaxies with direct hard
band detections (and the three sources with the smallest $\log f_{\rm
X}/f_{\rm R}$) are spectroscopically confirmed interlopers at $z<1$.
From Table~\ref{tab:ldist}, it is easy to see that much of the AGN
contamination in star-forming samples of galaxies (e.g, $\ugr$,
$\bzk$/SF) occurs for magnitudes $\ks<20$.

Figure~\ref{fig:xo} only shows those X-ray sources with hard band
detections.  Eleven additional sources had direct soft band
($0.5-2.0$~keV) detections, but no hard band detections.  Of these 11,
5 sources have $R>22.0$, $f_{\rm 0.5-2.0~keV} \la 0.1\times
10^{-15}$~erg~cm$^{-2}$~s$^{-1}$, and $\log f_{\rm X}/f_{\rm R} < -1$,
indicating they may be starburst galaxies.  These $5$ sources and
their properties are summarized in Table~\ref{tab:liksf}.  Three of
the five sources have spectra taken by us or by \citet{barger03}
indicating no obvious AGN spectral features.  These sources may be
rapidly star-forming galaxies and for fairness we include them in the
stacking analysis as indicated below and in Table~\ref{tab:ldist}.

\begin{figure}[hbt]
\centerline{\epsfxsize=8.5cm\epsffile{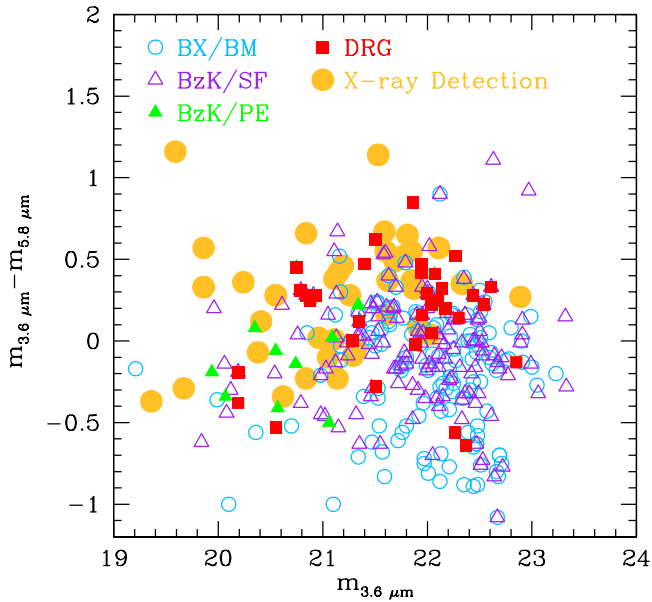}}
\figcaption[f6.eps]{{\it Spitzer}/IRAC $3.6-5.8$~$\mu$m color versus $3.6$~$\mu$m
  magnitude (in AB units) for all samples considered here, with
  emphasis on directly detected X-ray sources (large circles).  These
  direct detections generally have brighter IRAC magnitudes and redder
  colors than star forming $\ugr$ and $\bzk$/SF galaxies, likely due
  to thermal continuum from circumnuclear dust proximate to the AGN.
\label{fig:irac1m3}}
\end{figure}

It is interesting to also consider the rest-frame near-IR properties
of the directly detected X-ray sources as indicated by their {\it
Spitzer}/IRAC colors.  Figure~\ref{fig:irac1m3} shows the
$3.6-5.8$~$\mu$m color as a function of $3.6$~$\mu$m magnitude for all
samples considered here.  There is a clear segregation in the IRAC
colors of X-ray detections where they show, on average, brighter IRAC
magnitudes and redder IRAC colors when compared with the colors of
star-forming galaxies in the $\ugr$ and $\bzk$/SF samples.  Such a
trend might be expected if the rest-frame near-IR light from the X-ray
sources is dominated by thermal continuum from circumnuclear dust
heated by the AGN.  The increase in flux density across the IRAC bands
for AGN has been seen for ERO samples at redshifts $z\sim 1-3$
\citep{frayer04}, similar to what is observed here.  Finally, the $5$
objects listed in Table~\ref{tab:liksf} have $m_{\rm 3.6\mu m-5.8\mu
m} \sim -0.35$ to $0.35$ and $m_{\rm 3.6\mu m} = 20.2-22.0$, lying in
the same region of IRAC color space as some of the star-forming $\ugr$
and $\bzk$/SF candidates.

At times in the following analysis, we also consider submillimeter
galaxies and their relation to optical and near-IR selected objects.
These heavily star-forming objects are generally associated with
directly-detected X-ray sources, and we reconsider the X-ray emission
from these sources as pointed out below.  Unless otherwise stated,
however, we have excluded all directly-detected hard band X-ray
sources from the analysis under the assumption that their X-ray
emission is contaminated by AGN.

\begin{figure}[bht]
\centerline{\epsfxsize=8.5cm\epsffile{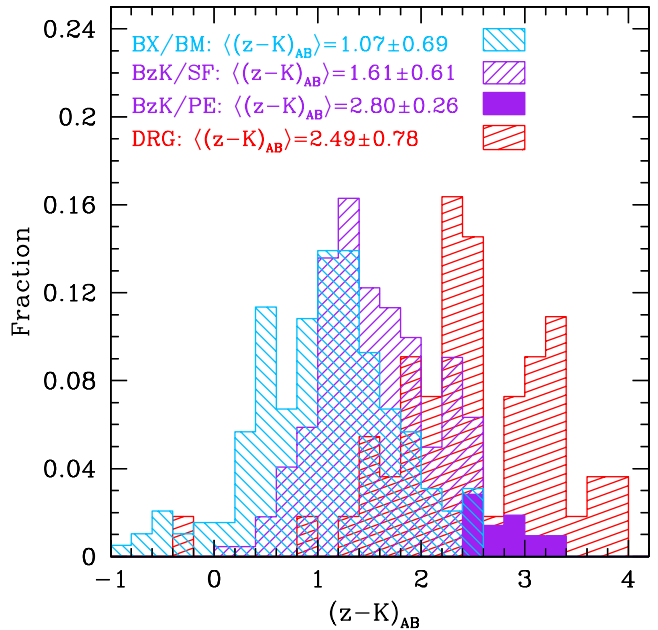}}
\figcaption[f7.eps]{$z-K$ color distribution for $\ugr$, star-forming $\bzk$, and
  DRG galaxies to $\ks=21$.  The mean $z-K$ color of galaxies becomes
  redder for the $\ugr$ to $\bzk$ to DRG samples.  $\ugr$ selection is
  more efficient in selecting objects with blue $\zmk\la 1$ and DRG
  selection is more efficient in selecting objects with very red
  $\zmk\ga 3$.  The $\bzk$ criteria spans the middle range of $\zmk$
  color.  The small solid histogram shows the arbitrarily normalized
  distribution in $\zmk$ color for passively-evolving $\bzk$ galaxies.
\label{fig:zmk}}
\end{figure}

\subsection{Overlap Between Samples}
\label{sec:zmk}

\begin{figure}[hbt]
\centerline{\epsfxsize=8.5cm\epsffile{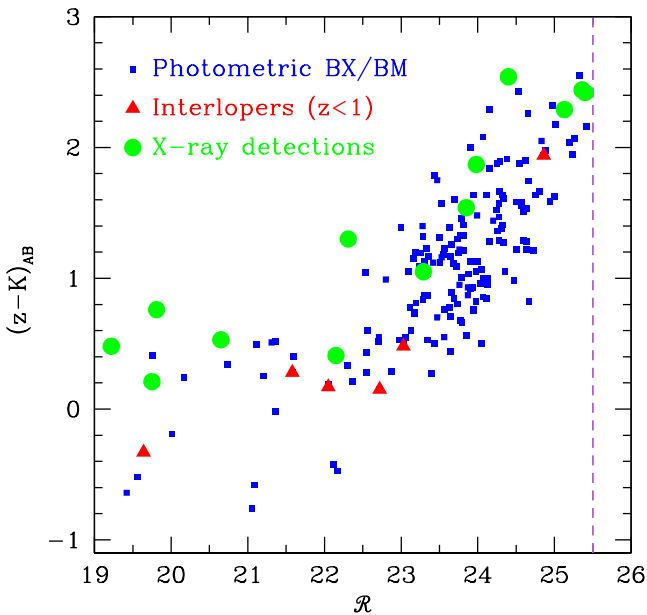}}
\figcaption[f8.eps]{$\zmk$ versus ${\cal R}$ for photometric $\ugr$ galaxies
  (squares) to $\ks=21$, showing that optically selected galaxies with
  red near-IR colors are optically fainter on average than those with
  blue near-IR colors.  This effect is due to the $\ks=21$ limit as
  well as the correlation between ${\cal R}$ and $z$-band magnitude as
  the two filters lie close in wavelength.  Objects with $\zmk \ga
  2.6$ are missed by $\ugr$ selection as they fall below the ${\cal
  R}=25.5$ $\ugr$ catalog limit (dashed vertical line).  Also shown
  are $\ugr$ sources with direct X-ray detections (large circles) and
  spectroscopically confirmed interlopers (triangles).
\label{fig:zmkvr}}
\end{figure}

Galaxies selected solely by the presence of some unobscured
star-formation ($\ugr$ selection), and those selected by some
combination of stellar mass and star-formation (DRG and $\bzk$
selection) can be distinguished by their observed near-IR color
distributions (Figure~\ref{fig:zmk}).  The mean $\zmk$ color for
$\ugr$ galaxies is $\sim 0.54$~mag bluer than the $\bzk$/SF sample,
just within the $1$~$\sigma$ dispersion of both samples.  This
difference in average $\zmk$ color between $\ugr$ and $\bzk$/SF
galaxies partly stems from the fact that the width of the $\bzk$
selection window below $\zmk=1$ narrows to the point where photometric
scatter becomes increasingly important in determining whether a galaxy
with blue colors (i.e., $\zmk<1$) is selectable with the $\bzk$/SF
criteria\footnote{None of the selection criteria considered here have
boxcar selection functions in either color or redshift space due to
various effects, including photometric errors.  The effect of this is
to suppress the efficiency for selecting objects whose intrinsic
colors lie close to the edges of the selection window.}.  On the other
hand, the $\ugr$ criteria are less efficient than $\bzk$ selection for
galaxies with $\zmk\ga1.6$.  $\ugr$ galaxies with red near-IR colors
are systematically fainter in the optical than those with blue near-IR
colors (Figure~\ref{fig:zmkvr}), reflecting both the correlation
between ${\cal R}$ and $z$ as these filters lie close in wavelength,
as well as the $\ks<21$ limit adopted in Figure~\ref{fig:zmkvr}.
Therefore, the optical catalog limit of ${\cal R}=25.5$ would appear
to exclude from the $\ugr$ sample those galaxies with $\zmk \ga 3$
(Figure~\ref{fig:zmkvr}).  As we show \S~\ref{sec:disc}, the exclusion
of $\zmk \ga 3$ galaxies by optical selection is not a fault of the
criteria themselves: the ${\cal R}=25.5$ limit is imposed so that
spectroscopic followup is feasible on the candidate galaxies.  Rather,
the exclusion of $\zmk\ga 3$ galaxies from optical surveys simply
reflects a fundamental change in the star formation properties of such
red galaxies.

\begin{figure}[hbt]
\centerline{\epsfxsize=8.5cm\epsffile{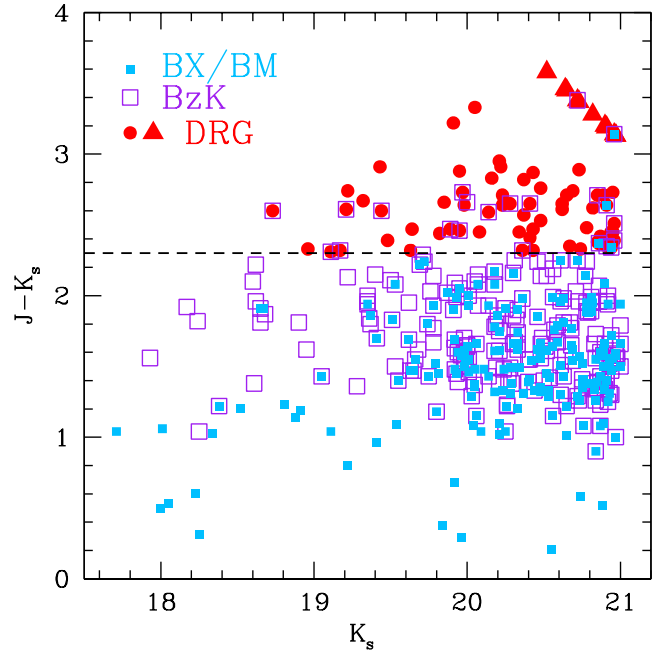}}
\figcaption[f9.eps]{$\jmk$ color versus $\ks$ for $\ugr$ (filled squares), $\bzk$
  (open squares), and DRG (circles) samples to $\ks=21$.  The hashed
  horizontal line denotes the $\jmk=2.3$ limit.  DRGs with limits in
  $J$-band are indicated by the triangles.  Approximately $5\%$ of
  DRGs satisfy the $\ugr$ criteria, but the fraction rises to $\sim
  12\%$ if we include those selected by the $z\sim 3$ LBG criteria.
\label{fig:jk}}
\end{figure}

\begin{figure}[hbt]
\centerline{\epsfxsize=8.5cm\epsffile{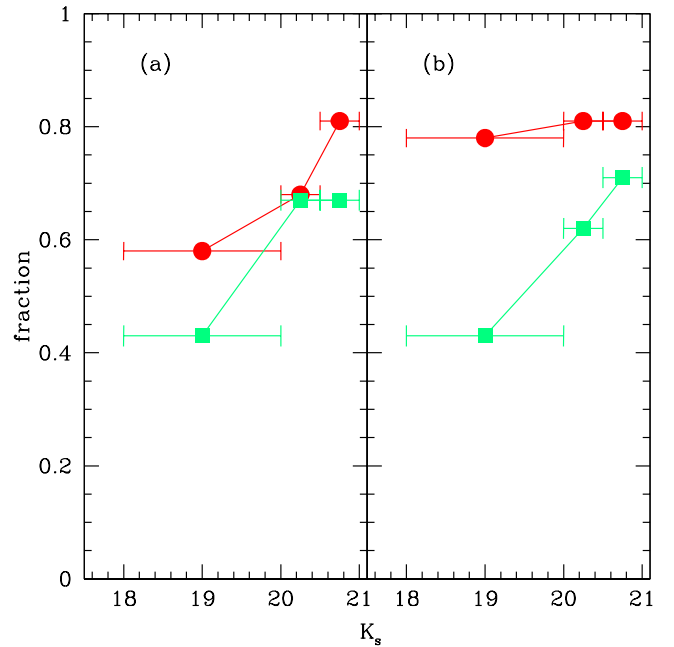}}
\figcaption[f10.eps]{(a) Fraction of $\bzk$/SF sources which are optically
  selected with the $\ugr$ criteria when including (squares) and
  excluding (circles) directly detected X-ray sources that are likely
  AGN (\S~\ref{sec:xray}); (b) Fraction of photometric BXs and BMs
  that are $\bzk$/SF selected (squares) and the fraction of BXs and
  BMs with confirmed redshifts $z>1.4$ that are $\bzk$/SF selected
  (circles).
\label{fig:frac}}
\end{figure}

\begin{figure}[thb]
\centerline{\epsfxsize=8.5cm\epsffile{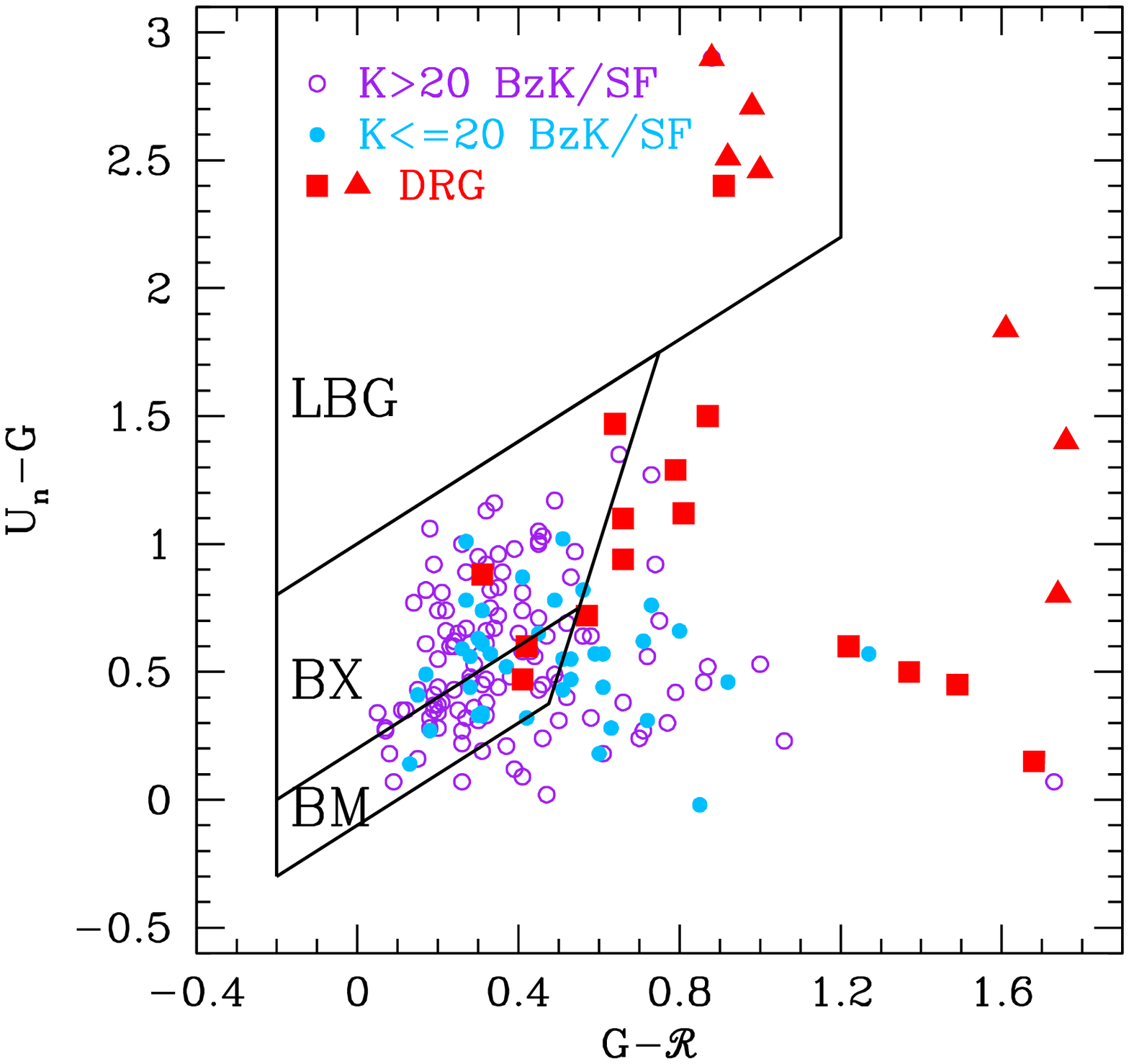}}
\figcaption[f11.eps]{$\ugr$ colors of $\bzk$/SF galaxies (circles) and DRGs with
  $\rs<25.5$, excluding direct X-ray detections.  Squares denote DRGs
  with $>5$~$\sigma$ detections in $U_{\rm n}$, $G$, and ${\cal R}$;
  triangles denote DRGs with $5$~$\sigma$ limits in $U_{\rm n}$.  All
  points are for galaxies with $\ks<21$.  Also indicated are the
  $\ugr$ selection criteria for $z\sim 2$ BMs and BXs, as well as the
  $z\sim 3$ LBG criteria of \citet{steidel03}.  Approximately $40\%$
  of our DRG sample galaxies have ${\cal R}<25.5$.  Of all DRGs,
  including those not shown in the figure and which have ${\cal
  R}>25.5$, $\sim 12\%$ can be selected using the BX, BM, or LBG
  criteria.  Note the number of $\bzk$ and DRG galaxies that lie very
  close (e.g., within $\la 0.2$~mag) of the $\ugr$ selection windows.
\label{fig:bzknotugr}}
\end{figure}

Separately, DRGs have a very red $\langle\zmk\rangle=2.49\pm0.78$.
Approximately $10\%$ of $z>1.4$ $\bzk$/SF galaxies also have
$\jmk>2.3$ (Figure~\ref{fig:jk}), similar to that found by
\citet{daddi04b}.  The fact that there is some, albeit small, overlap
between the $\bzk$/SF and DRG samples is not surprising since the two
criteria can be used to target reddened galaxies and both have redshift
distributions that overlap in the range $2.0<z<2.6$
(Figure~\ref{fig:zhist}).  The DRG fraction among $\bzk$ selected
galaxies does not change appreciably if we add in the $\bzk$/PE
sources---only $5$ of $17$ $\bzk$/PE galaxies have $\jmk>2.3$---as the
$\bzk$/PE galaxies are mostly at redshifts lower than the DRGs ($z\la
2$).  Finally, we note that DRGs include objects with much redder
$\zmk$ colors than found among $\ugr$ and $\bzk$/SF/PE galaxies,
i.e. those with $\zmk>3$.  The absence of these galaxies from
star-forming selected samples is discussed in \S~\ref{sec:pass}.

\begin{figure}[thb]
\centerline{\epsfxsize=8.5cm\epsffile{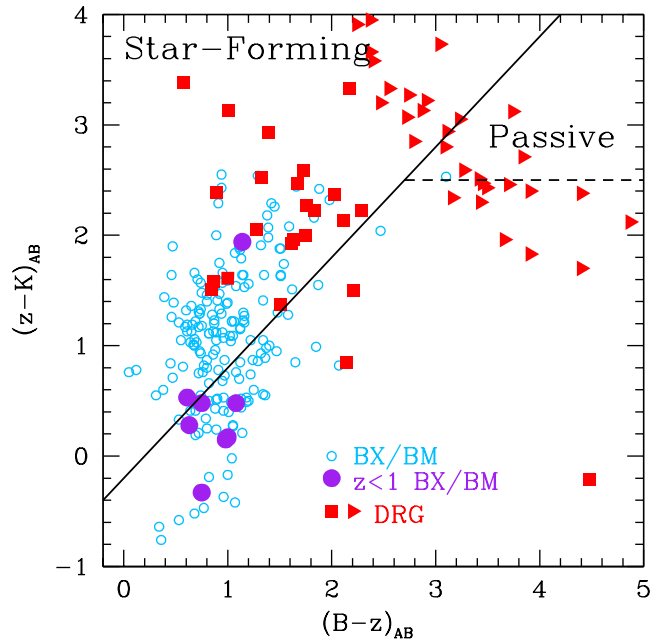}}
\figcaption[f12.eps]{$\bzk$ colors of $\ugr$ galaxies (empty circles) and DRGs
  (filled squares) to $\ks=21$.  Large filled circles denote $\ugr$
  objects with spectroscopically confirmed redshifts $z<1$
  (interlopers), most of which fall outside the $\bzk$/SF selection
  window.  Also shown is the expected region of color space for
  passively evolving $z>1.4$ galaxies \citep{daddi04b}.  Note the
  number of $\ugr$ galaxies that lie very close (e.g., within $\la
  0.2$~mag) of the $\bzk$ selection window.  DRGs with $B$-band
  limits, shown by the triangles, cluster in the region expected for
  passively evolving $z>1.4$ galaxies.
\label{fig:ugrnotbzk}}
\end{figure}

We can directly quantify the overlap between $\bzk$/SF and $\ugr$
galaxies.  Figure~\ref{fig:frac} shows the fraction of $\bzk$/SF
galaxies satisfying the $\ugr$ criteria (left panel) and the fraction
of $\ugr$ galaxies satisfying the $\bzk$/SF criteria (right panel).
Most of the contamination of the $\bzk$/SF sample (that we know of) is
from X-ray detected AGN (\S~\ref{sec:xray}), while most of the
contamination of the $\ugr$ sample is from low redshift interlopers
(Table~\ref{tab:kinterloper}).  Both sources of contamination tend to
occupy the bright end of the $K$-band apparent magnitude distribution.
We also show the overlap fractions in Figure~\ref{fig:frac} excluding
X-ray detected AGN and interlopers.  The $\ugr$ criteria recover an
increasing fraction of $\bzk$/SF selected sources proceeding from
$\ks<20$ galaxies ($\sim 60\%$ recovery fraction) to $\ks\sim 21$
galaxies ($\sim 80\%$ recovery fraction) after excluding directly
detected X-ray sources which are likely AGN (see \S~\ref{sec:xray}).
Conversely, the $\bzk$/SF criteria recover $\sim 80\%$ of
spectroscopically confirmed $\ugr$ galaxies at $z>1.4$, and are
evidently effective at recognizing most of the $\ugr$ low redshift
interlopers that tend to occupy the bright end of the $K$-band
apparent magnitude distribution.  This result stems from the fact that
low redshift interlopers tend to have bluer colors than necessary to
satisfy the $\bzk$/SF criteria.

Figures~\ref{fig:bzknotugr} and \ref{fig:ugrnotbzk} show that a
significant portion of $\bzk$/SF galaxies missed by $\ugr$ selection,
and conversely, have colors that place them within $\la 0.2$~mag of
the selection windows, which is comparable to the photometric
uncertainties.  The $\ugr$ criteria likely miss some $\bzk$/SF
galaxies not because of some failure of the criteria, but because we
cannot measure photometry with infinite precision.  The trend from
lower ($60\%$) to higher ($80\%$) recovery rate shown in
Figure~\ref{fig:frac}a reflects the fact that a greater percentage of
$\ks<20$ $\bzk$/SF galaxies have redder $G-{\cal R}$ colors (when
compared with $\ks>20$ $\bzk$/SF galaxies) than required to satisfy
the $\ugr$ criteria (Figure~\ref{fig:bzknotugr}).  There are some
$\bzk$/SF galaxies which have very red $G-{\cal R}\ga 0.8$ colors.  As
we show in \S~\ref{sec:disc}, these red $G-{\cal R}$ galaxies would
have an average bolometric SFR similar to $\bzk$/SF galaxies with
bluer $G-{\cal R}$ colors if they are at similar redshifts, $z\sim
2$.  Therefore, if these red objects are at $z\sim 2$, then
the correlation between $G-{\cal R}$ and reddening, as quantified by
the \citet{calzetti00} law, would appear to fail.  Photometric scatter
will also reduce the effectiveness of the $\bzk$ criteria in selecting
$\ugr$ galaxies (Figure~\ref{fig:ugrnotbzk}).  We can account for most
of the photometric incompleteness using the more sophisticated
analysis of \citet{reddy05}.

\begin{figure}[hbt]
\centerline{\epsfxsize=8.5cm\epsffile{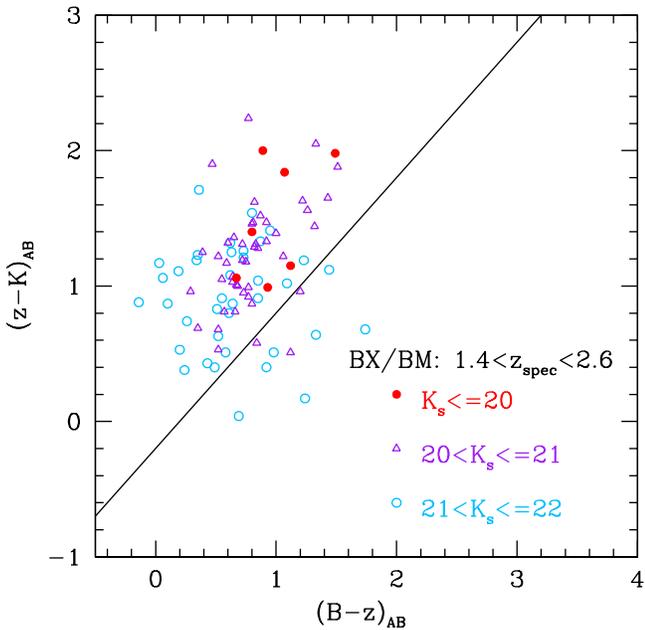}}
\figcaption[f13.eps]{$\bzk$ colors of spectroscopically confirmed $\ugr$ galaxies
  with redshifts $1.4<z<2.6$ for different bins in $\ks$ magnitude.
  $\bzk$/SF selection appears to miss an increasing fraction of
  $\ks>21$ galaxies in this redshift range due to the narrowing
  $\bzk$/SF selection window for objects with bluer $\zmk$ colors.  In
  addition, $\sim 19\%$ of $\ugr$ candidates have $\ks>22.0$ and are
  not shown.
\label{fig:faintk}}
\end{figure}

Our deep $K$-band data allow us to investigate the efficiency of
$\bzk$/SF selection to fainter $K$ magnitudes than previously
possible.  Figure~\ref{fig:faintk} shows the $\bzk$ colors of $\ugr$
galaxies with spectroscopic redshifts $1.4<z<2.6$ for three bins in
$\ks$ magnitude.  The $\bzk$/SF criteria were designed to select
relatively massive galaxies with $\ks<20$, but they become slightly
less efficient in culling $\ks>21$ galaxies: 10 of 49 ($\sim 20\%$)
$\ugr$ galaxies with spectroscopic redshifts $1.4<z<2.6$ and $\ks>21$
do not satisfy the $\bzk$/SF criteria.  Furthermore, we note that
$\sim 11\%$ ($61/544$) of $\ugr$ candidates that fall in the region
with $K$-band data are undetected to $\ks=22.5$ ($3$~$\sigma$).  The
$K$-band limits for these galaxies suggests they are younger
star-forming systems with $\zmk\la 1$, below which the $\bzk$/SF
criteria drop in efficiency, as discussed above.  We remind the reader
that many of the $\bzk$/SF objects not appearing in the $\ugr$ sample
may be missed by the $\ugr$ criteria simply because of photometric
errors.  $\ugr$ galaxies missed by the $\bzk$/SF criteria may be
missed not because of intrinsic differences in the objects, but simply
because of photometric scatter or because of the difficulty in
obtaining very deep $\ks$-band data.

Turning to $\jmk>2.3$ galaxies, we show the optical colors of DRGs
with ${\cal R}<25.5$ in Figure~\ref{fig:bzknotugr}, and the near-IR
colors of the $74\%$ of DRGs with $z$-band detections in
Figure~\ref{fig:ugrnotbzk}.  The optical criteria are particularly
inefficient in selecting $\jmk>2.3$ sources: $9$ of $73$ ($12\%$) of
DRGs in the GOODS-N field satisfy BX, BM, or LBG selection.  This
fraction is similar to the overall detection rate found by
\citet{erb05} for the $4$ fields in the $z\sim 2$ optical survey with
deep $J$ and $K$-band data.  The LBG criteria can be used to select
some DRGs since $z\sim 2$ galaxies with a Calzetti reddened constant
star formation SED with $\ebmv\ga 0.3$ are expected to lie in the
color space occupied by $z\sim 3$ LBGs.  A greater fraction ($\sim
30\%$) of DRGs satisfy the $\bzk$/SF criteria
(Figure~\ref{fig:ugrnotbzk}) since these criteria select objects with
developed spectral breaks and with redshifts that fall in the range
probed by DRG selection (see Figure~\ref{fig:zhist}).  We note that
DRGs with $B$-band limits cluster in the region of color space
expected for passively evolving $z>1.4$ galaxies
(Figure~\ref{fig:ugrnotbzk}).  These DRGs have little, if any, current
star formation (see \S~\ref{sec:pass}).

\subsection{Stacked X-ray Results}

The X-ray data are not sufficiently sensitive to detect individual
galaxies with SFR$\la 190$~$\sfr$ ($3$~$\sigma$).  We can, however,
stack the X-ray data for subsets of galaxies below the sensitivity
threshold to determine their average X-ray emission.  The influence of
AGN in any X-ray stacking analysis is a concern.  The softness of a
stacked signal provides some circumstantial evidence for X-ray
emission due primarily to star formation (e.g., \citealt{vandokkum04},
\citealt{daddi04b}, \citealt{laird05}).  UV line signatures and radio
emission can provide additional constraints on the presence of AGN
(e.g., \citealt{reddy04}).  We typically removed all directly-detected
X-ray sources from the optical and near-IR samples before running the
stacking simulations, except as noted below and in
Table~\ref{tab:liksf} when considering X-ray detected sources which
may be star-forming galaxies.  Our method of excluding other X-ray
detected sources ensures that luminous AGN do not contaminate the
stacked signal.  Indirect evidence suggests that less luminous AGN do
not contribute significantly to the stacked signal.  First, the
stacked signal has no hard band (HB; $2-8$~keV) detection indicating
that the signal is softer than one would expect with a significant AGN
contribution.  Second, the availability of rest-frame UV spectra for
many of the $\ugr$ objects provides an independent means of
identifying AGN.  There is one source whose spectrum shows
high-ionization emission lines in the rest-frame UV, but no X-ray
detection in the ${\it Chandra}$ $2$~Ms data.  Removing this X-ray
faint AGN source does not appreciably affect the stacked X-ray flux.
In addition, \citet{reddy04} examined the very same $\ugr$ dataset
used here and found a very good agreement between dust-corrected UV,
radio, and X-ray inferred SFRs for the sample, suggesting star
formation as the dominant mechanism in producing the observed
multi-wavelength emission.  Finally, the local hosts of low luminosity
AGN have stellar populations characteristic of passively evolving
early-type galaxies \citep{kauffmann03}.  In \S~\ref{sec:pass} we show
that passively evolving galaxies at $z\sim 2$ have little or no
detectable X-ray emission, implying that low level accretion activity
in these systems does little to alter the X-ray emission relative to
that produced from star formation.  The absence of X-ray emission from
these passively evolving galaxies also suggests that low mass X-ray
binaries contribute little X-ray emission in star-forming galaxies
when compared with the emission produced from more direct tracers of
the current star formation rate, such as high mass X-ray binaries.

\begin{figure}[hbt]
\centerline{\epsfxsize=8.5cm\epsffile{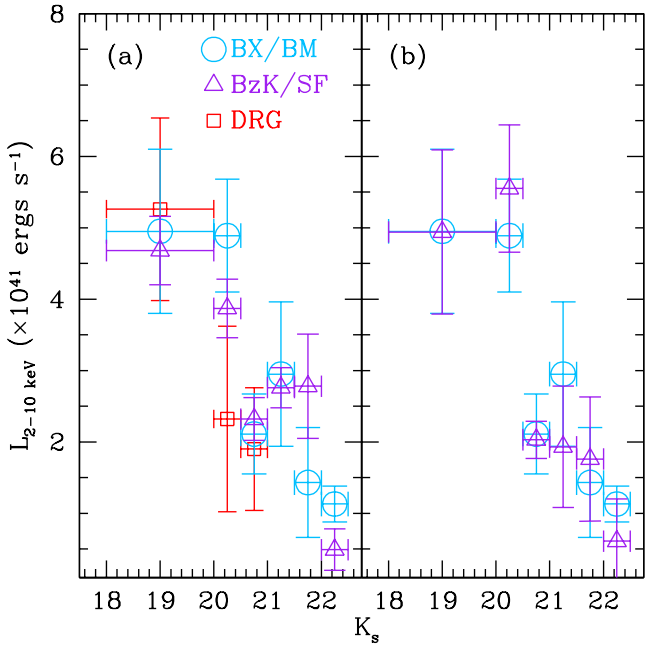}}
\figcaption[f14.eps]{Stacked X-ray luminosity versus $\ks$ magnitude for $\ugr$
  galaxies with $z>1$ (circles), $\bzk$/SF galaxies (triangles), and
  DRGs (squares), excluding all directly detected X-ray sources.  The
  left and right panels show the distributions for photometric and
  spectroscopically confirmed $z>1$ $\bzk$/SF galaxies, respectively.
  Sources without a spectroscopic redshift were assigned the mean
  redshift of the sample which they belong to, according to
  Figure~\ref{fig:zhist}.  In all cases, we find the distributions
  consistent within the errors.  We also find that $\ks<20$ galaxies
  have average X-ray luminosities that are a factor of $2-3$ times
  higher than that of $\ks>20.5$ galaxies.
\label{fig:ldist}}
\end{figure}

Stacking results for the samples (to $\ks=22.5$) are summarized in
Figure~\ref{fig:ldist} and Table~\ref{tab:ldist}.  The left panel of
Figure~\ref{fig:ldist} includes all photometrically-selected $\bzk$/SF
and DRG galaxies, and all spectroscopically-confirmed $z>1$ $\ugr$
galaxies. The right panel includes only those $\bzk$/SF galaxies with
spectroscopic redshifts $z>1$, all of which also satisfy the $\ugr$
criteria.  All direct X-ray detections have been excluded in making
Figure~\ref{fig:ldist}.  The distributions do not change appreciably
if we only consider the X-ray flux of $\bzk$/SF galaxies
spectroscopically confirmed to lie at $z>1$ (Figure~\ref{fig:ldist}b).
Removing the one spectroscopic $z>1$ AGN undetected in X-rays does
little to change the X-ray luminosity distributions.  The luminosity
distributions agree well between the three samples over a large range
in $\ks$ magnitude, with $\ks<20$ galaxies exhibiting the largest
X-ray luminosities by a factor of two to three when compared with
fainter $\ks>20.5$ galaxies.

\section{Discussion}
\label{sec:disc}

In this section we first present the X-ray inferred average bolometric
SFRs for galaxies in the $\ugr$, $\bzk$/SF/PE, and DRG samples, and
compare our results with other X-ray stacking analyses.  Unless stated
otherwise, we exclude hard band X-ray AGN sources from the analysis of
the SFRs.  The SFRs are interpreted for galaxies as a function of
their near-IR colors and we assess the ability of optical surveys to
single out both heavily reddened and massive galaxies.  We identify
passively evolving galaxies at $z\sim 2$ from their red near-IR colors
and discuss plausible star formation histories for these galaxies
using the X-ray data as an additional constraint.  Finally, we discuss
the contribution of $\ugr$, $\bzk$/SF, and DRG galaxies to the star
formation rate density at $z\sim 2$, taking into account the overlap
between the samples and their respective redshift distributions.

\subsection{Star Formation Rate Distributions}

\subsubsection{Star Formation Rates and Comparison with Other Studies}

\begin{figure}[hbt]
\centerline{\epsfxsize=8.5cm\epsffile{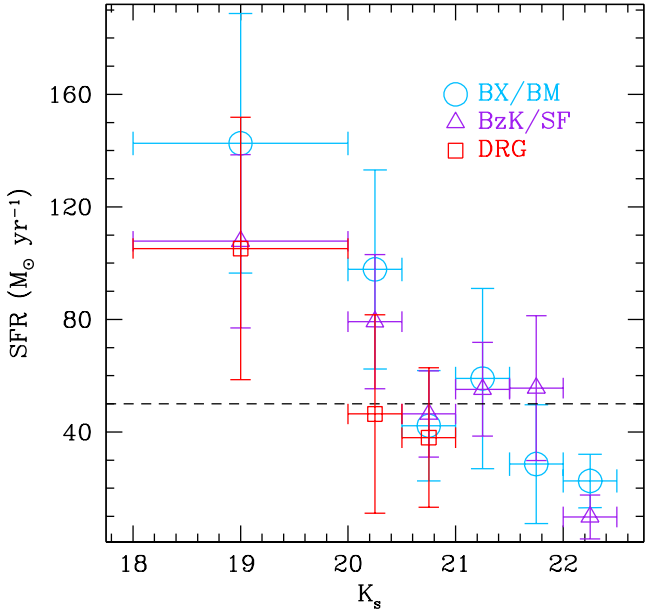}}
\figcaption[f15.eps]{Star formation rates inferred from the X-ray
  luminosity using the \citet{ranalli03} calibration.  We have added
  the $5$ directly-detected soft band X-ray sources that may be
  star-forming (Table~\ref{tab:liksf}) to compute the average SFR.
  Circles, triangles and squares denote $\ugr$, $\bzk$, and DRG
  samples, respectively.  The $\ugr$ points are for spectroscopically
  confirmed $z>1$ galaxies, and we have assumed the mean redshifts for
  the $\bzk$/SF and DRG samples as in Figure~\ref{fig:ldist}.
  Uncertainties in the star formation rates are dominated by scatter
  in the X-ray/FIR relation and the dispersion in the stacked X-ray
  estimates.  The dashed horizontal line denotes the average SFR for
  the entire spectroscopic ($z>1$) $\ugr$ sample of $\sim
  50$~M$_\odot$~yr$^{-1}$ from \citet{reddy04}.  Galaxies with
  $\ks<20$ have inferred star formation rates a factor of $2-3$ higher
  than for $\ks>20.5$ galaxies.
\label{fig:sfrdist}}
\end{figure}

We estimated the SFRs for galaxies in our samples using the
\citet{ranalli03} calibration between X-ray and FIR luminosity.  This
calibration reproduced the SFRs based on independent star formation
tracers for $z\sim 2$ galaxies \citep{reddy04}, so we are confident in
using it here.  The SFR distributions for $\ugr$, $\bzk$, and DRG
galaxies are shown in Figure~\ref{fig:sfrdist}, where we have added
the $5$ directly-detected soft band X-ray sources in
Table~\ref{tab:liksf} that may be star-forming galaxies.  The SFRs are
summarized in Table~\ref{tab:ldist}.  The mean SFR of $\ks<20$
galaxies is $\sim 90-140$~$\sfr$, and is a factor of $2-3$ times
larger than galaxies with $\ks>20.5$.  For comparison,
\citet{daddi04b} found an average SFR of K20 galaxies in the
GOODS-South field of $190$~$\sfr$ (including one likely star-forming
galaxy directly detected in X-rays).  This is somewhat higher than our
value of $110$~$\sfr$ for $\ks<20$ $\bzk$/SF galaxies.  This
discrepancy could simply result from field-to-field variations, small
number statistics, or the lower sensitivity of the X-ray data in the
GOODS-South field compared to GOODS-North.  With the {\it Chandra}
2~Ms data, we are able to exclude directly-detected X-ray sources down
to a factor of two lower threshold than was possible with the 1~Ms
data in the GOODS-South field.  If we add back those $\ks<20$ X-ray
$\bzk$/SF galaxies that would have been undetected in the 1~Ms data to
the stacking analysis, we obtain an average SFR of $160$~$\sfr$, more
in line with the \citet{daddi04b} value of $190$~$\sfr$.

A similar stacking analysis by \citet{rubin04} indicates that $\ks<22$
DRGs have SFRs of $\sim 280$~$\sfr$, corrected for the difference in
SFR calibration used here and in \citet{rubin04}.  This very high
value is likely a result of the shallow X-ray data ($74$~ksec)
considered in that study; the depth of their X-ray data precludes the
removal of most of the X-ray sources that are directly detected in the
2~Ms X-ray survey.  If we include those X-ray sources that would have
been undetected in the $74$~ksec data, assuming a mean redshift of
$\langle z\rangle=2.4$, the average SFR for the DRGs with $\ks<21$ is
$250$~$\sfr$.  Therefore, much of the difference in the SFRs can be
attributed to unidentified AGN in the shallower X-ray surveys
contaminating estimates of the star formation rate.  If
Figure~\ref{fig:sfrdist} is any indication, then adding DRGs with
$21<\ks\le22$ to the stack would decrease this average SFR.  Variance
of the fraction of DRG/PE galaxies between fields may also affect the
average SFRs: a greater fraction of DRG/PEs in the GOODS-N field,
$\sim 25\%$ (3/13) of which have $\ks<20$, will lead to a lower {\it
average} SFR for $\ks<20$ DRGs.  As we show in \S~4.4.2, there are
clearly some number of very reddened galaxies with large SFRs (e.g.,
submillimeter galaxies) among DRGs (and among the $\ugr$ and $\bzk$/SF
samples).  Regardless, these calculations underscore the importance of
factoring in the differing sensitivity limits of the various X-ray
surveys before comparing results.  The strong dependence of SFR with
$\ks$ magnitude (Figure~\ref{fig:sfrdist}) also suggests that fair
comparisons of the SFRs of galaxies selected in different surveys can
only be made between objects with similar rest-frame optical
luminosities.

Our analysis is advantageous as we are able to compare the SFRs of
galaxies within the same field, employing the same multi-wavelength
data (to the same sensitivity level) and the same photometric
measurement techniques, for a consistent comparison.  The inferred
average SFRs of $\bzk$/SF and DRG galaxies are remarkably similar to
that of optically selected galaxies once the samples are restricted to
similar $\ks$ magnitudes.  The previously noted discrepancies in X-ray
inferred SFRs of $\bzk$/SF, DRG, and $\ugr$ galaxies are therefore
likely a result of a mismatch between X-ray survey limits and near-IR
magnitude ranges.  Field-to-field variations may also partly account
for the previously observed discrepancies.

\subsubsection{Dependence of SFR on $\zmk$ Color}

\begin{figure}[hbt]
\centerline{\epsfxsize=8.5cm\epsffile{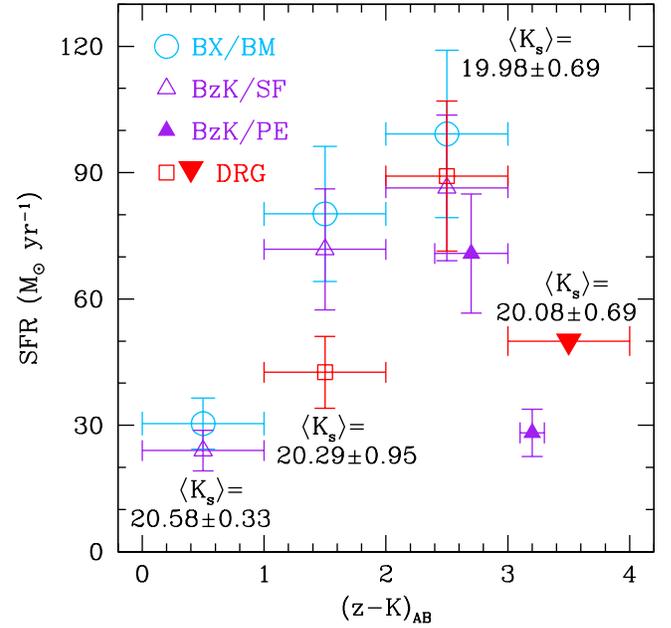}}
\figcaption[f16.eps]{Star formation rates of $\ks<21$ galaxies as a function of
  $\zmk$ color.  Symbols are the same as in Figure~\ref{fig:sfrdist}.
  The small solid triangles denote the average SFR inferred for $\bzk$
  galaxies that are selected to be passively evolving ($\bzk$/PE
  galaxies) and the large inverted triangle indicates the limit in SFR
  found for the $13$ DRGs with $\zmk>3.0$.  Note the turnover in
  inferred SFR at $\zmk \sim 3$.  Also indicated are the average $\ks$
  magnitudes for sources in each bin of $\zmk$ color.
\label{fig:sfrvzmk}}
\end{figure}

\begin{figure}[hbt]
\centerline{\epsfxsize=8.5cm\epsffile{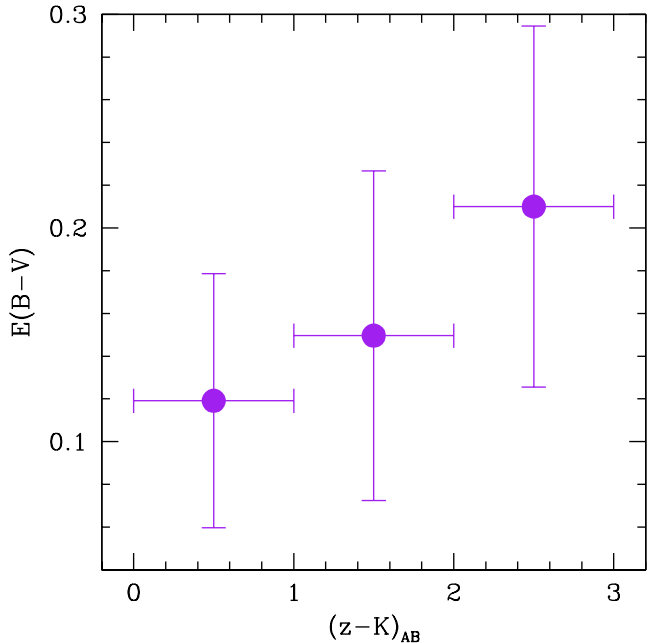}}
\figcaption[f17.eps]{Attenuation, as parameterized by the rest-frame UV spectral
  slope, $\ebmv$, as a function of $\zmk$ color for spectroscopically
  confirmed $\ugr$ galaxies with $z>1$.  Errors in $\ebmv$ represent
  the $1$~$\sigma$ dispersion of $\ebmv$ values for each bin of
  $\zmk$.
\label{fig:ebmvvzmk}}
\end{figure}

We began our analysis by noting the differences between the $\zmk$
color distributions of optical and near-IR selected galaxies
(Figure~\ref{fig:zmk}).  Figure~\ref{fig:sfrdist} indicates that
despite these near-IR color differences, the $\ugr$, $\bzk$, and DRG
galaxies have very similar {\it average} SFR distributions as a
function of $\ks$ magnitude.  Another proxy for stellar mass is the
$\zmk$ or $\rmk$ color (e.g., \citealt{shapley05}) as it directly
probes the strength of the Balmer and $4000$~\AA\, breaks.
Figure~\ref{fig:sfrvzmk} shows the inferred average SFRs of optical
and near-IR selected galaxies as a function of their $\zmk$ color,
excluding all directly-detected X-ray sources\footnote{Adding the $5$
sources in Table~\ref{tab:liksf} does not appreciably affect
Figure~\ref{fig:sfrvzmk}}.  Within any single sample, objects with red
$\zmk$ colors up to $\zmk\sim3$ have the largest SFRs.  The red $\zmk$
color for these objects with high SFRs likely results from a developed
spectral break (due to an older stellar population) combined with the
effects of dust.  In fact, Figure~\ref{fig:ebmvvzmk} illustrates the
tendency for $\ugr$ objects with spectroscopic redshifts $z>1$ and red
$\zmk$ colors to have larger attenuation, as parameterized by $\ebmv$,
on average, than those with bluer $\zmk$ colors.  The turnover in the
inferred SFR around $\zmk\sim 3$ is discussed in the next section.

Figure~\ref{fig:sfrvzmk} suggests that optically selected $\ugr$
galaxies may have systematically higher SFRs than $\bzk$ and DRG
galaxies with similar $\zmk$ colors, perhaps indicating that the
stacked sample for $\bzk$/SF and DRG galaxies includes those which are
passively evolving.  This may be particularly true of DRGs, quite a
few of which only have $B$-band limits and which cluster in the $\bzk$
color space occupied by passively evolving galaxies
(Figure~\ref{fig:ugrnotbzk}).  We can assess the dispersion in SFRs by
separately stacking galaxies that are expected to be currently
star-forming based on their colors and those that are not.  For
example, the average SFR computed for $\bzk$/SF galaxies that are {\it
not} selected by the $\ugr$ criteria (of which $\sim 70\%$ are within
$0.2$~mag of the $\ugr$ selection windows) is $\sim 70$~$\sfr$,
comparable to the average SFR of all $\bzk$/SF galaxies with $\ks<21$.
{\it In summary, star-forming $\bzk$ galaxies have similar SFRs
regardless of whether or not they satisfy the $\ugr$ criteria.}  The
$\ugr$ criteria miss some $\bzk$/SF galaxies either because of
photometric scatter or because their optical colors are not indicative
of their reddening.  It is worth noting that photometric scatter works
both ways: some sources with intrinsic colors satisfying the $\ugr$
criteria will be scattered {\it out} of the $\ugr$ selection windows
and some whose colors do not satisfy the $\ugr$ criteria will be
scattered {\it into} the $\ugr$ selection windows, although the two
effects may not equilibrate \citep{reddy05}.  The incompleteness of
the $\ugr$ and $\bzk$/SF criteria with respect to all star-forming
galaxies at $z\sim 2$ simply reflects our inability to establish
perfect selection criteria immune to the effects of photometric
scatter and SED variations while at the same time efficiently
excluding interlopers (e.g., \citealt{adel04}).  However, the
advantage of {\it spectroscopic} optical surveys is that their
selection functions can be quantified relatively easily (e.g.,
\citealt{adel04}, \citealt{reddy05}).

\subsubsection{Optical Selection of Reddened Star-Forming Galaxies}

Naively, one might interpret the inferred SFRs as a function of $\zmk$
color combined with the results shown in Figure~\ref{fig:zmkvr} to
suggest that $\ugr$ selection may miss the most actively star-forming
galaxies at $z\sim 2$.  We can interpret the similarity in SFRs of
$\ugr$ and $\bzk$/SF galaxies in the context of their reddening, as
parameterized by their rest-frame UV spectral slopes, $\ebmv$.
\citet{daddi04b} show that the reddening vector is essentially
parallel to the $\bzk$ limit defined by Equation~\ref{eq:bzkeq},
implying that $\bzk$ selection should be sensitive to galaxies with
higher extinction (and presumably higher star formation rates) than
found among $\ugr$ selected galaxies (i.e., $\ebmv\ge 0.3$).  However,
the similarity in the average SFRs of $\ugr$ and $\bzk$/SF galaxies
suggests several possibilities.  First, we noted above that $\bzk$/SF
galaxies not selected by the $\ugr$ criteria have similar SFRs as
those which do satisfy the $\ugr$ criteria\footnote{Those $\bzk$/SF
galaxies with $G-{\cal R}\ga 1$ and blue $U_{\rm n}-G\la 1$ have
optical colors that are similar to the colors expected for lower
redshift ($z\la 1$) galaxies (e.g., \citealt{adel04}).  So, if these
galaxies are truly low redshift galaxies, then their inferred star
formation rates would be even lower.}.  Consequently, $\bzk$/SF
galaxies with large SFRs that do not satisfy the $\ugr$ criteria
because they truly have $\ebmv>0.3$ may not exist in sufficient
numbers to significantly change the {\it average} SFRs for all
$\bzk$/SF galaxies which do not satisfy the $\ugr$ criteria.
\citet{adel00} and \citet{laird05} find that optically selected
galaxies with $z\ga 1$ show no correlation between their rest-frame UV
luminosities and their obscuration, implying that on average the
redder (more obscured) galaxies have higher bolometric SFRs than
galaxies with less reddening.  Therefore, the similarity in the
average X-ray inferred SFRs of $\ugr$ and $\bzk$/SF galaxies suggests
that there are not large numbers of galaxies with $\ebmv\ga 0.4$
(i.e., if there were a large number of such heavily reddened objects,
their bolometric SFRs would imply X-ray luminosities large enough to
be directly detected in the soft band X-ray data, and very few likely
star-forming galaxies at these redshifts are directly detected in the
soft band).

\begin{deluxetable*}{lcccccccccc}[tbh]
\tabletypesize{\footnotesize}
\tablewidth{0pc}
\tablecaption{Properties of Submillimeter Galaxies with $\ks$-band Data}
\tablehead{
\colhead{$\alpha$\tablenotemark{a}} & 
\colhead{$\delta$\tablenotemark{a}} &
\colhead{$S_{\rm 850\mu m}$\tablenotemark{b}} & 
\colhead{$\ks$} &
\colhead{} &
\colhead{$f_{\rm SB} \times 10^{-15}$\tablenotemark{d}} &
\colhead{$f_{\rm HB} \times 10^{-15}$\tablenotemark{d}} &
\colhead{} &
\colhead{} & 
\colhead{} & 
\colhead{$L_{\rm bol}$\tablenotemark{f}} \\
\colhead{(2000.0)} &
\colhead{(2000.0)} &
\colhead{(mJy)} &
\colhead{(Vega mag)} &
\colhead{z\tablenotemark{c}} &
\colhead{(erg~s$^{-1}$~cm$^{-2}$)} &
\colhead{(erg~s$^{-1}$~cm$^{-2}$)} &
\colhead{$\ugr$\tablenotemark{e}} &
\colhead{$\bzk$/SF\tablenotemark{e}} &
\colhead{DRG\tablenotemark{e}} &
\colhead{($\times 10^{12}$~L$_{\odot}$)}}
\startdata
12:36:21.27 & 62:17:08.4 & $7.8\pm1.9$ & 20.62 & 1.988 & ... & ...   & yes & yes & yes & 13.5 \\ 
12:36:22.65 & 62:16:29.7 & $7.7\pm1.3$ & 19.85 & 2.466 & ... & 1.14  & yes & yes & no  & 11.6 \\
12:36:29.13 & 62:10:45.8 & $5.0\pm1.3$ & 17.65 & 1.013 & 0.17 & 2.23 & no  & no  & no  & 1.2 \\ 
12:36:35.59 & 62:14:24.1 & $5.5\pm1.4$ & 18.62 & 2.005 & 0.21 & 2.48 & no  & yes & no  & 8.1 \\
12:36:36.75 & 62:11:56.1 & $7.0\pm2.1$ & 18.41 & 0.557 & 1.62 & 2.01 & no  & no  & no  & 0.12 \\
12:36:51.76 & 62:12:21.3 & $4.6\pm0.8$ & 18.34 & 0.298 & 0.34 & 2.65 & no  & no  & no  & 0.08 \\ 
12:37:07.21 & 62:14:08.1 & $4.7\pm1.5$ & 20.05 & 2.484 & 0.09 & 0.91 & no  & no  & yes & 7.5 \\ 
12:37:12.05 & 62:12:12.3 & $8.0\pm1.8$ & 20.65 & 2.914 & 0.03 & 0.37 & no  & no  & yes & 5.5 \\ 
12:37:21.87 & 62:10:35.3 & $12.0\pm3.9$ & 17.59 & 0.979 & 0.05 & 2.11 & no  & no  & no  & 0.53 \\
\\
12:37:11.98 & 62:13:25.7 & $4.2\pm1.4$ & 20.61 & 1.992 & ... & 1.01 & yes & yes & no & 4.9 \\
12:37:11.34 & 62:13:31.0 & $4.4\pm1.4$ & 18.65 & 1.996\tablenotemark{g} & 0.07 & 0.52 & no & yes & no & ... \\
\enddata
\tablenotetext{a}{Radio coordinates are from Table~2 of \citet{chapman05}.  Five of these sources
are also in \citet{wang04}.  The last source listed is only from \citet{wang04}.}
\tablenotetext{b}{$S_{\rm 850\mu m}$ fluxes are from Table~2 of \citet{chapman05}.  Two of the sources
are measured by \citet{chapman05} to have $S_{\rm 850\mu m}\sim 4.6-4.7$~mJy, and are measured by 
\citet{borys03} and \citet{wang04} to have $S_{\rm 850\mu m}>5$~mJy, and for fairness we include these in the table.}
\tablenotetext{c}{Spectroscopic redshift from \citet{chapman05}.}
\tablenotetext{d}{Soft and hard band fluxes are from \citet{alexander03}.}
\tablenotetext{e}{This field indicates whether the submillimeter source satisfies
the $\ugr$, $\bzk$/SF, and DRG selection criteria.}
\tablenotetext{f}{Inferred bolometric luminosity from \citet{chapman05}.  The mean
bolometric luminosity of the 6 submillimeter galaxies with spectroscopic redshifts
$1.4<z<2.6$ is $\langle L_{\rm bol}\rangle\sim 9\times 10^{12}$~L$_{\odot}$.}
\tablenotetext{g}{Redshift from \citet{swinbank04}.}
\label{tab:smg}
\end{deluxetable*}

Second, studies of the UV emission from submillimeter galaxies (SMGs)
suggest that heavily reddened galaxies may have similar rest-frame UV
spectral properties, such as their range in $\ebmv$, as those which
are forming stars at modest rates, implying that the correlation
between $\ebmv$ and bolometric SFR (e.g., from the \citealt{meurer99}
and \citealt{calzetti00} laws) breaks down for the most actively
star-forming galaxies (e.g., \citealt{chapman05}).
Table~\ref{tab:smg} summarizes the properties of the $9$ {\it known}
radio-selected SMGs with $S_{\rm 850\mu m}\ga 5$~mJy in the GOODS-N
field that overlap with our near-IR imaging
(\citealt{chapman05};\citealt{wang04}).  Also listed are two (pair) sources
with $S_{\rm 850\mu m} \sim 4$~mJy taken from \citet{chapman05} and
\citet{wang04}.  Of the 7 SMGs with redshifts $1.4<z<2.6$, 3 satisfy
the $\ugr$ criteria.  The detection rate of $\sim 40\%$ is similar to
the detection rate of SMGs with $\ugr$ colors found by
\citet{chapman05}.  The mean bolometric luminosity of the $5$ SMGs is
$\langle L_{\rm bol}\rangle\sim 9\times 10^{12}$~L$_{\odot}$ as
inferred from their submillimeter emission, corresponding to an SFR of
$\sim 1500$~$\sfr$ using the \citet{kennicutt98} relation\footnote{As
we discuss in \S~\ref{sec:sfrd}, some of the submillimeter flux may be
coming from accretion activity}.  Despite their large bolometric
luminosities, the three submillimeter galaxies with redshifts in our
sample have dust-corrected UV SFRs of $14-28$~$\sfr$.  In these cases,
the UV emission may come from a relatively unobscured part of the
galaxy or may be scattered out of the optically-thick dusty regions
\citep{chapman05}.  The $\bzk$/SF criteria cull 5 of the 7 SMGs with
redshifts $1.4<z<2.6$.  Therefore, at least in the small sample of
SMGs examined here (irrespective of their X-ray properties), the
$\ugr$ and $\bzk$/SF samples host an approximately equal number of
SMGs.  Finally, as we show below, galaxies with the most extreme
$\zmk$ colors (i.e., $\zmk>3$) are red not because they are obscured
by dust, but because they have little or no current star formation.
It is therefore not surprising that such objects are not identified by
criteria designed to select star-forming galaxies.

\subsection{Passively Evolving Galaxies at $z\sim 2$}
\label{sec:pass}

\subsubsection{Near-IR Colors}

We now turn to galaxies in our samples that appear to have little or
no current star formation.  DRGs have SFRs that are comparable to
those of $\bzk$/SF and $\ugr$ galaxies with similar near-IR colors for
$\zmk<3$.  However, stacking the $13$ DRGs with $\zmk\ge 3$ results in
a non-detection with an upper limit of $\sim 50$~$\sfr$
(Figure~\ref{fig:sfrvzmk}).  Stacking the X-ray emission from the $17$
$\bzk$/PE galaxies shows a similar turnover in the inferred average
SFR around $\zmk\sim 3$ (Figure~\ref{fig:sfrvzmk}).  The $\bzk$ colors
of the DRGs (most of which only have $B$-band limits) lie in the
$\bzk$ color space expected for passively-evolving galaxies
(Figure~\ref{fig:ugrnotbzk}).  The stacking analysis confirms that
these red DRGs and $\bzk$/PE galaxies have little current star
formation compared with DRGs and $\bzk$/PE galaxies with bluer near-IR
colors (Figure~\ref{fig:sfrvzmk}).  A similar X-ray stacking analysis
by \citet{brusa02} yields no detection for passive EROs in the K20
survey from the {\it Chandra} Deep Field South data.

The average $\jmk$ color of the $13$ passively evolving DRGs (with
$\zmk>3$) is $\langle\jmk\rangle=2.98\pm0.59$, and is comparable to
the average $\jmk$ color of DRGs with bluer $\zmk$ colors, implying
that the $J$-band is either in close proximity to the Balmer and
$4000$~\AA\, breaks or the band encompasses the breaks completely.
This will occur for galaxies with a mean redshift $\langle
z\rangle\sim 1.88-2.38$ (Figure~\ref{fig:sed}).  In these cases, the
$\zmk$ color will be more effective than the $\jmk$ color in culling
those galaxies with developed spectral breaks.  The fraction of
passively evolving DRGs inferred from their lack of X-ray emission is
$13/54\sim24\%$, which is in reasonable agreement to the passively
evolving DRG fraction of $\sim 30\%$ found by \citet{forster04} and
\citet{labbe05}.

Alternatively, galaxies with $2<\zmk<3$ must still be forming stars at
a prodigious rate, as indicated by their stacked X-ray flux.  The
similarity in average $\ks$ magnitude between galaxies with $2\le\zmk
< 3$ and $\zmk\ge 3$ (Figure~\ref{fig:sfrvzmk}) suggests they have
similar masses, and the difference in $\zmk$ color between the two
samples simply reflects the presence of some relatively unobscured
star formation in those galaxies with $2\le\zmk<3$.

\subsubsection{Stellar Populations}

\begin{figure*}[tbh]
\plottwo{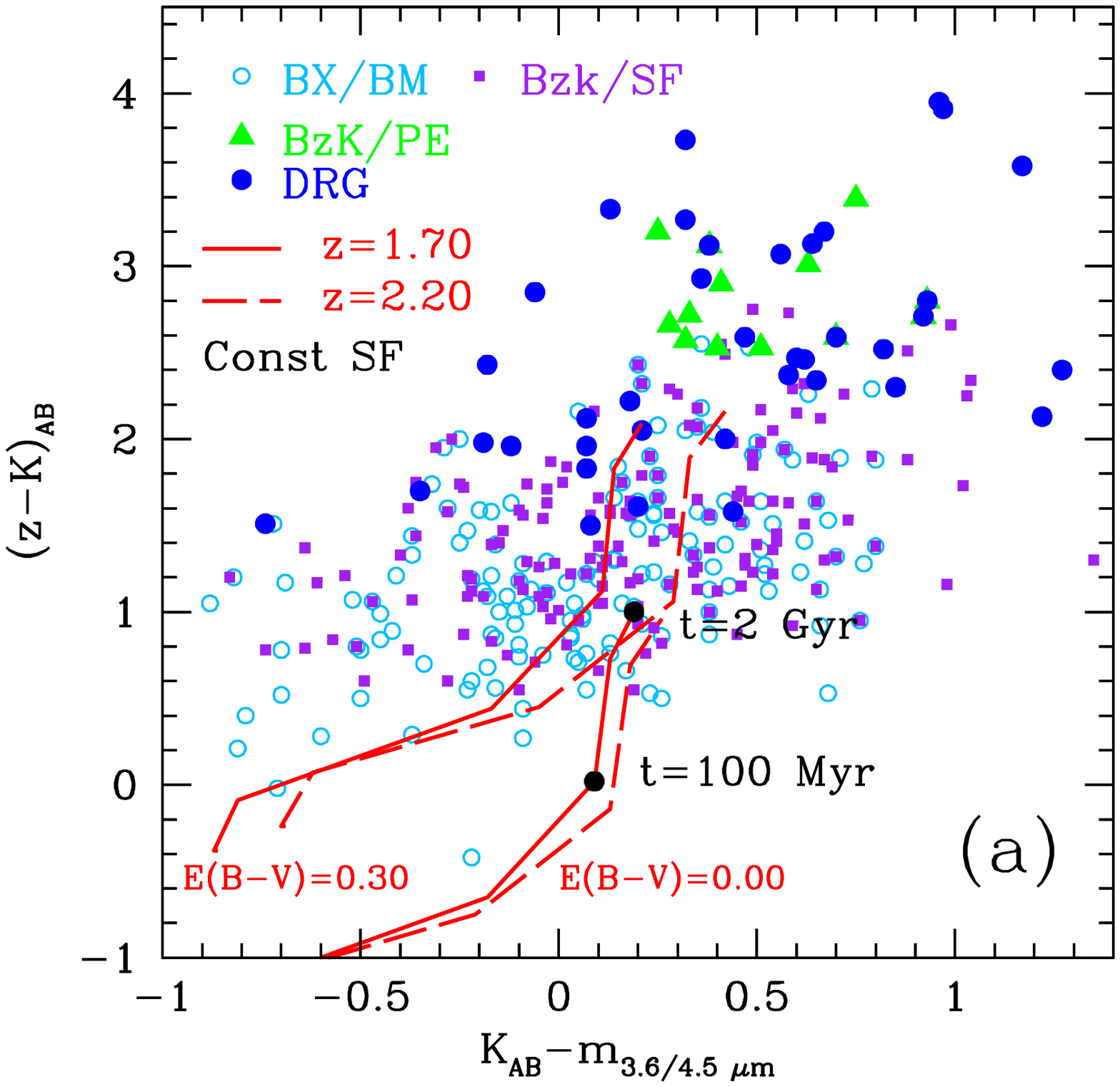}{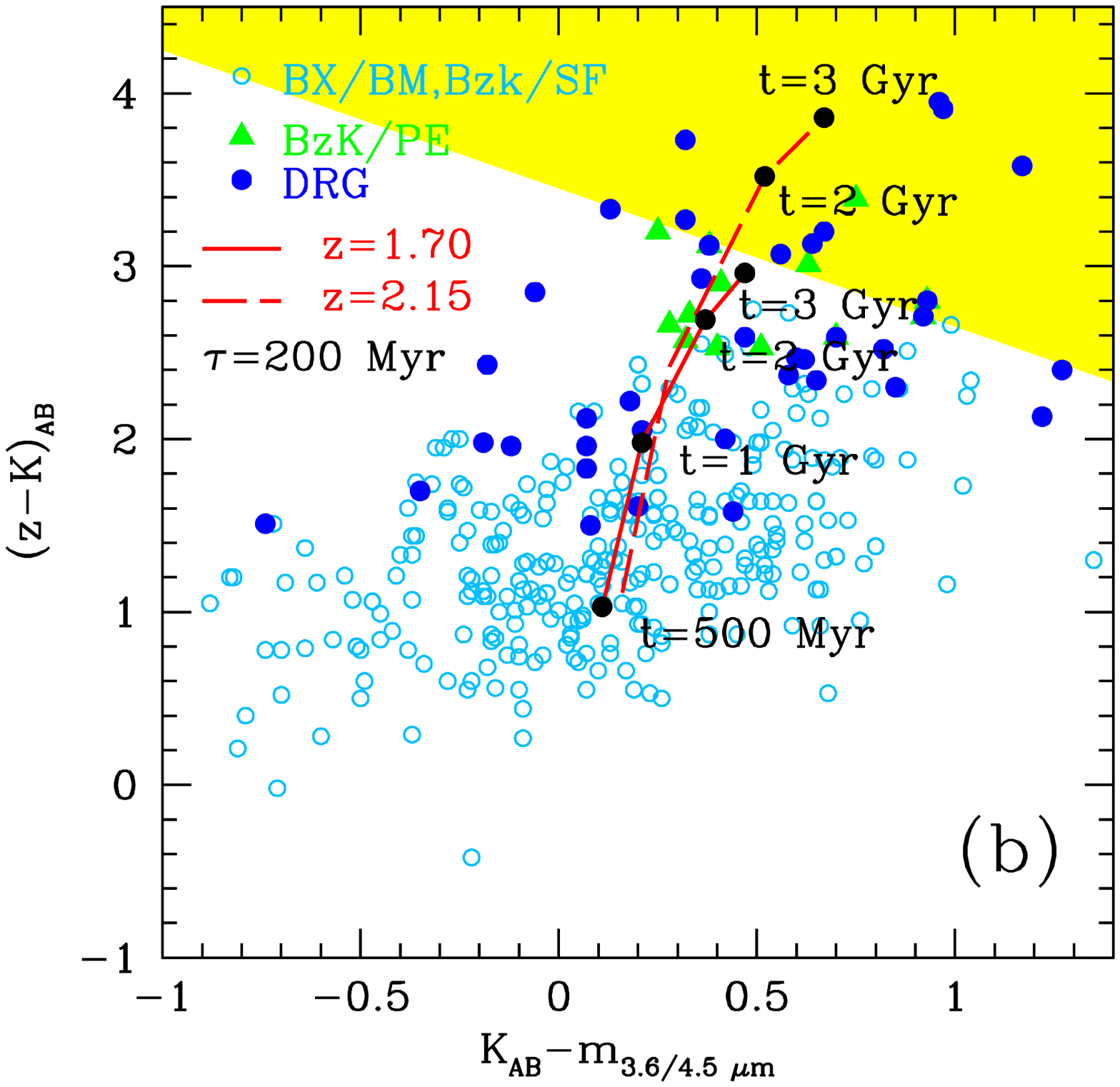}
\caption{(a) $\zmk$ versus $(K-m_{\rm 3.6/4.5\mu m})_{\rm AB}$ colors
  for $\ks<21$ galaxies in the $\ugr$, $\bzk$/SF, $\bzk$/PE, and DRG
  samples.  Also shown are \citet{bruzual03} spectral templates for
  ages 1 Myr to 3 Gyr at the mean redshifts of the BX ($\langle
  z\rangle \sim 2.2$) and BM ($\langle z \rangle \sim 1.7$) samples,
  assuming constant star formation, $\ebmv=0.0$ and $0.3$, and the
  \citet{calzetti00} reddening law.  The models assume a Salpeter IMF
  from $0.1-100$~M$_{\odot}$ and solar metallicity. The photometric
  scatter is large given that we include galaxies with IRAC
  uncertainties up to $0.5$~mag, and this accounts for the large
  spread in the $(K-m_{\rm 3.6/4.5\mu m})_{\rm AB}$ colors of the
  star-forming candidates; (b) same as (a) except we show the spectral
  templates for a model with $\tau=200$~Myr and $\ebmv = 0.0$.  The
  shaded region selects IRAC Extremely Red Objects (IEROs;
  \citealt{yan04}).
\label{fig:synth}}
\end{figure*}

Figure~\ref{fig:synth} further illustrates the differences between the
star-forming and passively evolving galaxies in terms of some physical
models.  The left panel of Figure~\ref{fig:synth} shows the $\zmk$
versus $(K-m_{\rm 3.6/4.5\mu m})_{\rm AB}$ colors (near-IR/IRAC color
diagram) for all galaxies, excluding direct X-ray detections
\footnote{\citet{labbe05} propose a similar diagram to differentiate
DRGs from other (e.g., star-forming) populations of $z\sim 2$
galaxies.}.  Because the SEDs of (non-AGN) galaxies considered here
are expected to be relatively flat in $f_{\nu}$ across the IRAC bands,
we used $3.6$~$\mu$m IRAC AB magnitudes for all sources that were not
covered by the $4.5$~$\mu$m imaging.  Also shown in
Figure~\ref{fig:synth}a are synthetic colors for \citet{bruzual03}
spectral templates at the mean redshifts of the BM ($\langle
z\rangle\sim 1.7$) and BX ($\langle z\rangle\sim 2.2$) samples,
assuming constant star formation, $\ebmv=0$ and $0.3$, and a
\citet{calzetti00} reddening law.  The bulk of $\ugr$ and $\bzk$/SF
galaxies generally fall within the region of color space expected for
constant star forming galaxies with moderate extinction of $\ebmv\sim
0.15$ and ages of $\sim 1$~Gyr.  These values are consistent with
those derived from detailed spectral modeling of $\ugr$ galaxies and
LBGs by \citet{shapley05}.  Much of the scatter of star-forming
galaxies to the left and right of the CSF models for $\ebmv=0.15-0.30$
is a result of photometric uncertainty, particularly in the $(K-m_{\rm
3.6/4.5\mu m})_{\rm AB}$ color, since we include galaxies with formal
IRAC uncertainties of $0.5$~mag.  In addition, some of the scatter for
objects with blue $\zmk$ colors arises from interlopers.  The more
interesting aspect of Figure~\ref{fig:synth}a is that the constant
star formation models cannot account for the colors of objects with
$\zmk\ga 3$: these objects must have ages less than the age of the
universe at $z\sim 2$ ($\sim 3$~Gyr) and simultaneously have modest
$\ebmv$---and hence modest current SFRs$\la 190$~$\sfr$---such that
they remain undetected as soft-band X-ray sources.  The important
result is that, similar to SED fitting, the X-ray stacking analysis
allows us to rule out certain star formation histories.  For the PE
galaxies considered here, the X-ray data rule out the constant star
formation models.  The benefit of X-ray data is that we can quantify
the current SFR independent of extinction and the degeneracies which
plague SED fitting.

The only models that can account for the colors of objects with
$\zmk\ga 3$ are those with declining star formation histories (right
panel of Figure~\ref{fig:synth}).  For example, DRG/PEs at $z\sim 2.2$
have colors that can be reproduced by dust free models ($\ebmv=0.0$)
with star formation decay timescales between $\tau=10$~Myr
(instantaneous burst) and $\tau\sim700$~Myr\footnote{We rule out ages
that are greater than the age of the universe at $z\sim 2.2$, giving
an upper limit on the age of $\sim 3$~Gyr.}.  While the upper limit on
the current average SFRs of DRG/PEs of $50$~$\sfr$
(Figure~\ref{fig:sfrvzmk}) does not help us further constrain the star
formation history to a narrower range in $\tau$, the fact that DRG/PEs
are still detected at $z$-band suggests that single stellar population
models with small $\tau$ are unrealistic (e.g., ages greater than
$1$~Gyr imply $>100$ e-folding times for the instantaneous burst model
making such an object undetected at $z$-band).  If there is ongoing
low level star formation activity, then a two component model with a
underlying old stellar population and a recent star formation episode
may be required (e.g., \citealt{yan04}).

The red $\zmk>3$ colors of $\bzk$/PE galaxies can be reproduced by
models with $\tau\sim 10-300$~Myr (for $\tau$ much larger than
$300$~Myr, the models over predict the current star formation rate).
The X-ray data indicate that these $\bzk$/PE galaxies have an average
current SFR of $\sim 28$~$\sfr$.  For an age of $1$~Gyr and
$\tau=300$~Myr, this implies an ``initial'' SFR of $\sim 800$~$\sfr$
at $z\sim 2.4$.  This initial SFR is comparable to that of SMGs and
the implied formation redshift is close to the median redshift of SMGs
($z\sim 2.2$), suggesting SMGs could be plausible progenitors of
$\bzk$/PE galaxies if the single component model correctly described
the star formation histories of these galaxies (e.g.,
\citealt{cimatti04}).  The formation redshifts can be pushed back in
time to significantly earlier epochs $z\ga 3.5$ if one assumes a more
physically motivated ``truncated'' star formation history that models
the effects of feedback in halting star formation
\citep{daddi05}.  Nonetheless, the simplistic example above
illustrates how X-ray estimates of the bolometric SFRs of galaxies can
be combined with the results of stellar population modeling to
indicate the likely progenitors of such galaxies.  In summary, we have
demonstrated the utility of stacked X-ray data as a powerful
constraint on the results from stellar population modeling.  The X-ray
data indicate that galaxies with $\zmk\ga 3$ have SEDs that are
consistent with declining star formation history models.  Other
studies also show that the SEDs of $\bzk$/PE galaxies can be
adequately described by declining star formation history models (e.g.,
\citealt{daddi05}).  The deep X-ray data confirm these results and
further allow us to constrain the current SFRs of passively evolving
and star-forming galaxies independent of the degeneracies associated
with stellar population modeling.

\subsubsection{Space Densities}

To conclude this section, we note that objects selected by their
$\jmk$ colors appear to include a substantial number of passively
evolving galaxies at redshifts $z\ga 2$ (although, as we pointed out
above, the $\zmk$ color may be a more effective means of determining
{\it which} DRGs are passively evolving).  The space density implied
by the $4$ $\bzk$/PE galaxies with $\zmk>3$ and $\ks<21$ is $\sim
3\times 10^{-5}$~Mpc$^{-3}$, assuming a boxcar (or top-hat) selection
function between redshifts $1.4<z<2.0$ and an area of $\sim
72.3$~arcmin$^{2}$.  If we include all $17$ $\bzk$/PE galaxies (i.e.,
including those with $\zmk<3$ and those four that are
directly-detected in X-rays) with $\ks<21$, we find a space density of
$1.3\times 10^{-4}$~Mpc$^{-3}$.  Given the strong clustering observing
for $\bzk$/PE galaxies (e.g., \citealt{daddi05}), our estimate is in
good agreement with the value of $1.8\times 10^{-4}$~Mpc$^{-3}$
obtained for the \citet{daddi05} sample of {\it bona-fide} $\bzk$/PE
objects in the Hubble Ultra Deep Field (UDF), an area which is $6$
times smaller than the area considered in our analysis (i.e.,
$12.2$~arcmin$^{2}$ in \citet{daddi05} versus $72.3$~arcmin$^{2}$
considered here).

All of the $\bzk$/PE galaxies would have been detected with $\ks<21$
and $\zmk>3$, assuming the PE model shown in Figure~\ref{fig:synth}b,
if they were at the mean redshift assumed for DRG/PEs of $\langle
z\rangle =2.2$.  The space density of the $13$ DRG/PEs with $\zmk>3$
is $\sim 9\times 10^{-5}$ to $\ks=21$, assuming a boxcar selection
function between between redshifts $2.0<z<2.6$ and an area of $\sim
72.3$~arcmin$^{2}$.  Here, we have assumed that the range in redshifts
of DRG/PEs ($2.0<z<2.6$) is similar to that of all DRG galaxies based
on the spectroscopic redshift distribution shown in
Figure~\ref{fig:zhist}.  Our estimate of the DRG/PE space density is
comparable to that obtained by \citet{labbe05} after restricting their
sample to $\ks<21$ (yielding $1$ object over $\sim 5$~arcmin$^{2}$),
and assuming a volume between redshifts $2.0<z<2.6$).  

It is worth noting that the $\bzk$/PE and DRG/PE populations appear to
be highly clustered (e.g., \citealt{daddi05}, \citealt{vandokkum04})
and this will likely affect their density estimates over small
volumes.  While the space densities derived here are in rough
agreement with other studies, our estimates have been derived over a
much larger volume (by a factor of $6-14$) than any previous study and
will be less susceptible to variations in density due to clustering.
We caution the reader that the density estimates may still be
uncertain given that the redshift distribution of DRGs with $z\ga 2.6$
is not well sampled, even in the large {\it spectroscopic} dataset of
DRGs considered here.

Taken at face value, the density estimates derived above suggest a
significant presence of passively evolving $\ks<21$ galaxies at
redshifts $z\ga 2$.  This result contrasts with that of
\citet{daddi05} who argue for a rapid decrease in the number density
of passively evolving galaxies at redshifts $z\ga 2$ based on
$\bzk$/PE selection.  The dropoff in space density of passively
evolving galaxies at redshifts $z\ga 2$, as suggested by
\citet{daddi05}, may be an artifact of the $\bzk$/PE selection
function which, based on previously published redshift distributions
and shown in Figure~\ref{fig:zhist}, appears to miss passively
evolving galaxies at redshifts $z\ga 2$, even when using the very deep
{\it B}-band of this study.  Figure~\ref{fig:ugrnotbzk} shows that all
of the DRGs that cluster around the $\bzk$/PE selection window have
limits in $B$-band and very few actually have limits that would for
certain place them in the $\bzk$/PE window.  Photometric scatter for
those galaxies with very faint (or no) $B$-band detections is likely
to be significant for these galaxies.  These results suggest that the
depth of the $B$-band data is the determining factor in whether
$\bzk$/PE selection culls galaxies with redshifts $z\ga 2$ or not.
Because the depth of the photometry is an issue for the $\bzk$/PE
selection, it becomes difficult to accurately quantifying with a
single selection criteria the dropoff in space density of passively
evolving galaxies between $z\la 2$ and $z>2$.  We can avoid the need
for excessively deep $B$-band data to select passively evolving
galaxies with redshifts $z\ga 2$ by simply selecting them using a
single color, $\jmk$ or $\zmk$.  The stacked X-ray results show that a
subsample of DRGs have very little star formation, suggesting that
passively evolving galaxies have a significant presence at epochs
earlier than $z=2$.  The inferred ages of DRGs would imply formation
redshifts of $z\sim 5$ \citep{labbe05}.

We conclude this section by noting that there are several galaxies in
the HS1700+643 sample of \citet{shapley05}, and many more in optically
selected samples in general (e.g., \citealt{erb05}), which have old
ages and early formation redshifts similar to those of the passively
evolving $\bzk$/PE and DRG/PE galaxies discussed here.  In order to
reproduce the observed SEDs for such objects, the current SFR must be
much smaller (but still detectable in the case of optically selected
galaxies) than the past average SFR.

\subsection{Selecting Massive Galaxies}

As discussed above, DRGs with $\zmk>3$ appear to be passively evolving
based on their (lack of) stacked X-ray flux and their colors with
respect to models with declining star formation histories.  The X-ray
data indicate that $\bzk$/PE galaxies also appear to be well described
by declining star formation histories, consistent with the SED
modeling results of \citet{daddi05}.  The stellar mass estimates of
these PE galaxies will be presented elsewhere.  Here, we simply
mention that several existing studies of the stellar populations of
$\bzk$/PE and DRG galaxies with $\ks<20$ indicate they have masses
$\ga 10^{11}$~M$_{\odot}$ (e.g., \citealt{daddi05},
\citealt{forster04}).  In addition, \citet{yan04} recently analyzed
the stellar populations of IRAC-selected Extremely Red Objects
(IEROs), selected to have $f_{\nu}({\rm 3.6\mu m})/ f_{\nu}({\rm
z_{\rm 850}}) > 20$ (or, equivalently, $(z-{\rm 3.6\mu m})_{\rm AB} >
3.25$).  Spectral modeling indicates these sources lie at redshifts
$1.6<z<2.9$, are relatively old ($1.5-3.5$~Gyr), and require an
evolved stellar population to fit the observed SEDs.  Almost all of
the PE galaxies with $\zmk>3$ satisfy the IERO criteria (shaded region
of Figure~\ref{fig:synth}b).  Furthermore, the ${\cal R}$-band
detections and limits for PEs with $\zmk>3$ imply $\rmk\ga 5.3$,
satisfying the ERO criteria.  

\begin{figure}[thb]
\centerline{\epsfxsize=8.5cm\epsffile{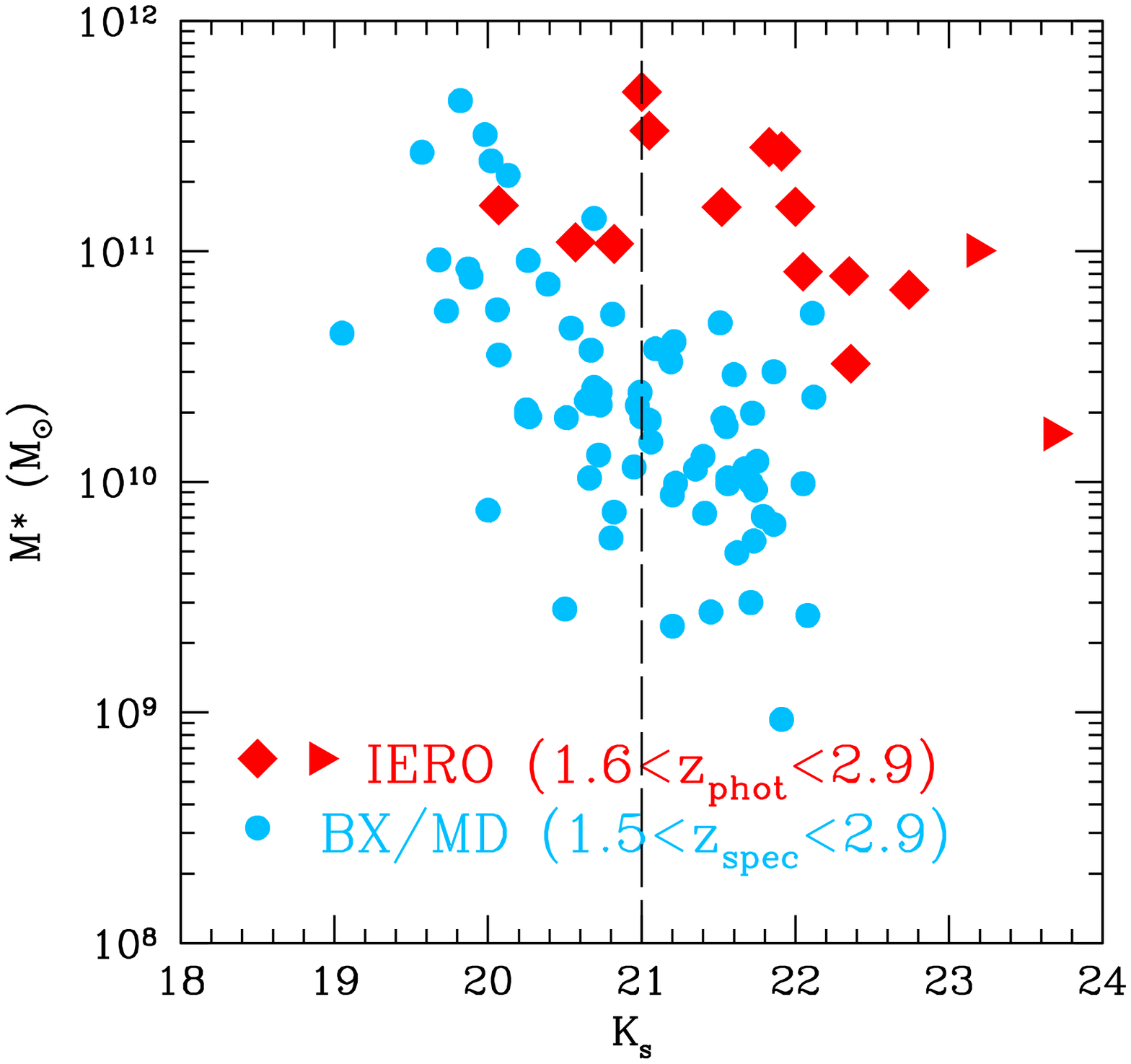}}
\figcaption[f19.eps]{Inferred stellar masses of BX and ``MD'' objects from the
  \citet{shapley05} sample with spectroscopic redshifts $1.5<z_{\rm
  spec}<2.9$ (circles), and IEROs from the \citet{yan04} sample with
  photometric redshifts $1.6<z_{\rm phot}<2.9$ (squares, and triangles
  for those with $\ks$-band limits).  The stellar masses from
  \citet{yan04} have been multiplied by 1.7 to convert from a Chabrier
  to Salpeter IMF.  The dashed vertical line denotes the limit
  brighter than which we are complete for IEROs (i.e., those DRGs with
  $\zmk>3$).  A subset of BX galaxies have stellar masses similar to
  those of the IERO sample.  The scatter in stellar masses reflects at
  least a magnitude variation in the $M/L$ ratio of BX/MD and IERO
  objects at a given rest-frame optical luminosity.
\label{fig:iero}}
\end{figure}

The inferred stellar masses of the \citet{shapley05} sample of
optically-selected galaxies in the HS1700+643 field are shown in
Figure~\ref{fig:iero}.  For comparison, we also show the inferred
stellar masses from the \citet{yan04} sample of IEROs\footnote{H. Yan,
private communication}.  Both the \citet{shapley05} and \citet{yan04}
samples take advantage of the longer wavelength IRAC data to constrain
the stellar masses, and the typical uncertainty in mass is $\sim 40\%$
for objects with $M^*<10^{11}$~M$_{\odot}$ and $\le 20\%$ for objects
with $M^*>10^{11}$~M$_{\odot}$.  The IERO stellar masses have been
multiplied by $1.7$ to convert from a Chabrier to Salpeter IMF.  The
scatter in Figure~\ref{fig:iero} reflects the large ($\ga
1$~magnitude) variation in the mass-to-light ($M/L$) ratio for objects
with a given rest-frame optical luminosity in the BX and IERO samples.
For magnitudes brighter than our DRG completeness limit of $\ks=21$,
$\sim 16\%$ of BX galaxies have stellar masses $M^* >
10^{11}$~M$_{\odot}$.  For $\ks<20$ BX galaxies, the fraction with
$M^* > 10^{11}$~M$_{\odot}$ is $\sim 40\%$.  So, while
optically-selected BX galaxies have a lower {\it mean} stellar mass
than IEROs, there is certainly a subsample of BX galaxies which have
masses comparable to the most massive IERO galaxies.  We note that the
stellar mass distributions of $\ks>21$ BXs and IEROs do not overlap:
the $\rmk$ colors of IEROs are too red for them to be included in the
optically-selected sample.

The X-ray stacking results indicate that BX/BM galaxies with $\ks<20$
have prodigious star formation rates, while IEROs (i.e., those DRGs
with $\zmk>3$) have very little star formation.  Therefore, a simple
interpretation is that optical surveys include objects that are as
massive ($M^*>10^{11}$~M$_{\odot}$) as those selected in near-IR
surveys, with the only requirement that the galaxies have some
unobscured star formation.  The range of star formation rates
(uncorrected for extinction) found for $\ugr$ galaxies is
$3-60$~$\sfr$, and it is likely that massive galaxies with at least a
little unobscured star formation can be $\ugr$ selected.  This may be
the only significant difference between optical and near-IR selected
massive galaxies, and the difference in SFR may be temporal.  These
criteria typically fail to select passively evolving galaxies at
$z\sim 2$ as they have already settled to a quiescent stage.  This
does not mean that such massive galaxies will never appear in optical
surveys.  For example, a subsequent accretion event at $z<2$ could
elevate the star formation activity in an otherwise passively evolving
massive galaxy, thus bringing it into the optical sample.
Nonetheless, the DRG and $\bzk$/PE criteria add to the census of
galaxies at $z\sim 2$ by selecting passively evolving galaxies that
have stellar masses similar to the most massive galaxies selected in
the rest-frame UV.

\begin{deluxetable*}{llccc}[tbh]
\tabletypesize{\footnotesize}
\tablewidth{0pc}
\tablecaption{Cumulative Contributions to the SFRD Between $1.4<z<2.6$}
\tablehead{
\colhead{} &
\colhead{} &
\colhead{} & 
\colhead{$\rho$\tablenotemark{c}} &
\colhead{SFRD\tablenotemark{d}} \\
\colhead{$\ks$ Range} &
\colhead{Sample\tablenotemark{a}} &
\colhead{$z$\tablenotemark{b}} &
\colhead{(arcmin$^{-2}$)} &
\colhead{($\sfr$~Mpc$^{-3}$)}}
\startdata
$\ks\leq 20.0$ & $\ugr$ & 1.4--2.6 & 0.44 & $0.016\pm0.008$ \\
               & $\bzk$/SF & 1.4--2.6 & 0.62 & $0.018\pm0.004$ \\
               & DRG/SF & 2.0--2.6 & 0.15 & $0.008\pm0.002$ \\
               & $\bzk$/SF---$\ugr$ & 1.4--2.6 & 0.35 & $0.007\pm0.002$ \\
               & DRG/SF---$\bzk$/SF---$\ugr$ & 2.0--2.6 & 0.08 & $0.004\pm0.001$ \\
               & {\bf Total} & 2.0--2.6 & 0.87 & $0.027 \pm 0.006$ \\ 
\\
$\ks\leq 20.5$ & $\ugr$ & 1.4--2.6 & 1.13 & $0.034\pm0.009$ \\
               & $\bzk$/SF & 1.4--2.6 & 1.35 & $0.032\pm0.006$ \\
               & DRG/SF & 2.0--2.6 & 0.30 & $0.013\pm0.003$ \\
               & $\bzk$/SF---$\ugr$ & 1.4--2.6 & 0.55 & $0.012\pm0.002$ \\
               & DRG/SF---$\bzk$/SF---$\ugr$ & 2.0--2.6 & 0.25 & $0.009\pm0.002$ \\
               & {\bf Total} & 2.0--2.6 & 1.93 & $0.055 \pm 0.011$ \\
\\
$\ks\leq 21.0$ & $\ugr$ & 1.4--2.6 & 2.15 & $0.045\pm0.010$ \\
               & $\bzk$/SF & 1.4--2.6 & 2.34 & $0.044\pm0.009$ \\
               & DRG/SF & 2.0--2.6 & 0.58 & $0.020\pm0.004$ \\
               & $\bzk$/SF---$\ugr$ & 1.4--2.6 & 0.74 & $0.014\pm0.002$ \\
               & DRG/SF---$\bzk$/SF---$\ugr$ & 2.0--2.6 & 0.41 & $0.013\pm0.002$ \\
               & {\bf Total} & 2.0--2.6 & 3.30 & $0.072\pm 0.014$ \\
\\
$\ks\leq 21.5$ & $\ugr$ & 1.4--2.6 & 3.17 & $0.057\pm0.013$ \\
               & $\bzk$/SF & 1.4--2.6 & 3.61 & $0.052\pm0.010$ \\
               & $\bzk$/SF---$\ugr$ & 1.4--2.6 & 1.71 & $0.017\pm0.002$ \\
               & {\bf Total}\tablenotemark{e} & 2.0--2.6 & 5.29 & $0.087\pm 0.017$ \\
\\
$\ks\leq 22.0$ & $\ugr$ & 1.4--2.6 & 4.31 & $0.069\pm0.015$ \\
               & $\bzk$/SF & 1.4--2.6 & 4.93 & $0.068\pm0.014$ \\
               & $\bzk$/SF---$\ugr$ & 1.4--2.6 & 2.55 & $0.020\pm0.003$ \\
               & {\bf Total}\tablenotemark{e} & 2.0--2.6 & 7.27 & $0.102\pm 0.021$ \\
\enddata
\tablenotetext{a}{The $\bzk$/SF---$\ugr$ sample represents the set of objects that are
$\bzk$/SF-selected, but not $\ugr$-selected.  Similarly, the DRG/SF---$\bzk$/SF---$\ugr$
sample represents the set of DRGs that are not selected by either the $\bzk$/SF or $\ugr$
criteria.}
\tablenotetext{b}{Redshift range of sample.}
\tablenotetext{c}{Surface density of photometric objects after removing interlopers and directly
detected X-ray sources that are likely AGN.  The number of $\ugr$ objects is
calculated assuming the spectroscopic and interloper fractions from Table~\ref{tab:kinterloper}.  
The overlap fractions are taken from Figure~\ref{fig:frac}.  We assume a field area of 
$\sim 72.3$~arcmin$^{2}$ to compute surface densities.}
\tablenotetext{d}{Assumes the average SFRs shown in Figure~\ref{fig:sfrdist}.}
\tablenotetext{e}{This includes the contribution from the DRG/SF---$\bzk$/SF---$\ugr$ sample for
$\ks<21$.}
\label{tab:sfrdtab}
\end{deluxetable*}

\subsection{Star Formation Rate Density at $z\sim 2$}
\label{sec:sfrd}

\subsubsection{Contribution from Optical and Near-IR Selected Samples}

\begin{figure}[hbt]
\centerline{\epsfxsize=8.5cm\epsffile{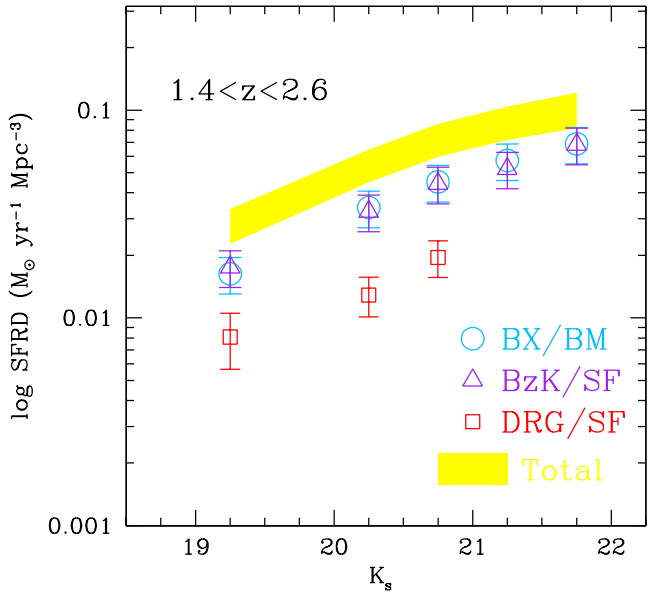}}
\figcaption[f20.eps]{{\it Cumulative} star formation rate density (SFRD) as a
  function of $\ks$ magnitude for $\ugr$ (circles), $\bzk$/SF
  (triangles), and DRG/SF (squares) galaxies with redshifts
  $1.4<z<2.6$.  The points are not independent of each other given the
  overlap between the samples.  The shaded region denotes the total
  cumulative SFRD when counting overlap objects once.  The total SFRD
  to $\ks=22$ includes DRGs with $\ks<21$.  The error bars reflect the
  Poisson error and uncertainty in star formation rate added in
  quadrature, but do not reflect systematic errors associated with,
  e.g., photometric scattering.
\label{fig:sfrd}}
\end{figure}

We can roughly estimate the contribution of $\ugr$, $\bzk$/SF, and
DRG/SF galaxies to the extinction-free star formation rate density
(SFRD) at $z\sim 2$\footnote{Although the $\bzk$/PE galaxies do have
detectable X-ray emission (e.g., Figure~\ref{fig:sfrvzmk}), their
contribution to the SFRD is minimal given that their space density is
a factor of 5 smaller than that of the $\bzk$/SF and $\ugr$ galaxies
to $\ks=21$ in the redshift range $1.4<z<2.0$.}.  The $\bzk$/SF
criteria are designed to select galaxies with redshifts $1.4<z<2.6$.
The similarity in surface densities, volumes probed, and SFRs of
galaxies in the $\bzk$/SF and $\ugr$ samples implies that their
contribution to the SFRD will be comparable for objects with $\ks<22$.
The redshift distribution of $\ks<21$ DRGs from within our own sample
is reasonably well defined over this redshift range (cf.,
Figure~\ref{fig:zhist}), so we can estimate the added contribution of
$\sim 80\%$ of the DRGs with redshifts $2.0\la z<2.6$ to the SFRD
between redshifts $1.4<z<2.6$.  Figure~\ref{fig:sfrd} and
Table~\ref{tab:sfrdtab} show the cumulative contribution to the SFRD
of $\bzk$/SF, $\ugr$, and DRG/SF galaxies.  The points in
Figure~\ref{fig:sfrd} are not independent of each other due to the
overlap between the samples (e.g., Figure~\ref{fig:frac}).  Also shown
in Figure~\ref{fig:sfrd} by the shaded region is the inferred total
SFRD assuming the overlap fractions of Figure~\ref{fig:frac} and
counting all objects once.  The results indicate that $\ugr$ selection
would miss approximately one-third of the total SFRD from $\ugr$ and
$\bzk$/SF galaxies to $\ks=22$ and DRG/SF $\ks<21$ galaxies combined.
We remind the reader that much of the incompleteness of the $\ugr$
sample with respect to that of the $\bzk$/SF sample (and vice versa)
results from photometric scattering (e.g., Figures~\ref{fig:bzknotugr}
and \ref{fig:ugrnotbzk}).  Monte Carlo simulations can be used to
quantify the biases of such photometric inaccuracy and thus correct
for incompleteness (e.g., \citealt{reddy05}).  The total SFRD in the
interval $1.4<z<2.6$ for $\ugr$ and $\bzk$/SF galaxies to $\ks=22$ and
DRG/SF galaxies to $\ks=21$, taking into account the overlap between
the samples, is $\sim 0.10\pm0.02$~$\sfr$~Mpc$^{-3}$.  Approximately
$30\%$ of this comes from galaxies with $\ks<20$
(Table~\ref{tab:sfrdtab}).  Optically-selected galaxies to ${\cal
R}=25.5$ and $\ks=22.0$ and $\bzk$/SF galaxies to $\ks=22.0$ (with
significant overlap between the two samples) account for $\sim 87\%$
of the total SFRD quoted above.  DRGs to $\ks=21$ that are not
selected by the $\ugr$ or $\bzk$/SF criteria contribute the remaining
$\sim 13\%$.  We note that the number $\sim
0.10\pm0.02$~$\sfr$~Mpc$^{-3}$ does not include the $5$ radio-selected
SMGs to $S_{\rm 850\mu m}\ga 5$~mJy that are near-IR and/or
optically-selected since we removed the directly-detected hard-band
X-ray sources in computing the SFRD.  If we add these $5$
radio-selected SMGs that are present in the optical and near-IR
samples (all of which have $\ks<21$), then the total SFRD contributed
by the $\ugr$ and $\bzk$/SF objects to $\ks=22.0$ and DRG/SF galaxies
to $\ks=21.0$ is $\sim 0.15\pm 0.03$~$\sfr$~Mpc$^{-3}$ (see next
section).

\subsubsection{Contribution from Radio-Selected Submillimeter Galaxies}

We conclude this section by briefly considering the contribution of
radio-selected submillimeter galaxies (SMGs) with $S_{\rm 850\mu m}\ga
5$~mJy to the SFRD.  All but one of the radio-selected SMGs summarized
in Table~\ref{tab:smg} are directly detected in either the soft or
hard band and are likely associated with AGN.  Stacking the X-ray
emission for the $5$ radio-selected SMGs with redshifts $1.4<z<2.6$ in
Table~\ref{tab:smg} yields an average inferred SFR of $\sim
2900$~$\sfr$, and this value should be regarded as an upper limit
given that the X-ray emission is likely contaminated by AGN.  On the
other hand, the average bolometric luminosity of the $5$ SMGs, as
derived from their submillimeter flux, is $\langle L_{\rm
bol}\rangle\sim 9\times 10^{12}$~L$_{\odot}$.  If we assume that
$30\%$ of $L_{\rm bol}$ arises from AGN (e.g., \citealt{chapman05},
\citealt{alexander05}), then the implied SFR is $\sim 1000$~$\sfr$.
If we take at face value the assertion that $30\%$ of the bolometric
luminosity of submillimeter galaxies comes from AGN, then this means
the X-ray emission would overestimate the average SFR of SMGs by a
factor of $\sim 3$.  In other words, only one-third of the X-ray
emission from SMGs would result from star-formation, and the remaining
two-thirds would come from AGN.

To determine the {\it additional} SFRD provided by radio-selected SMGs
with $S_{\rm 850\mu m}\ga 5$~mJy, we must account for their overlap
with the optical and near-IR samples.  The data in Table~\ref{tab:smg}
show that there are 4 of 9 SMGs that are not selected by the optical
and/or near-IR criteria.  All 4 of these galaxies have relatively low
redshifts $z\la 1$ and will obviously not contribute to the SFRD
between redshifts $1.4<z<2.6$.  Alternatively, of the $5$ SMGs that
are spectroscopically confirmed to lie at redshifts $1.4<z<2.6$, {\it
all} are selected by either the $\ugr$, $\bzk$/SF, or DRG criteria
(and sometimes by more than one set of criteria).

Because of the non-uniform coverage of the submillimeter observations,
we must rely on the published submillimeter number counts to estimate
the effective surface density probed by the 9 radio-selected SMGs with
$S_{\rm 850\mu m}\ga 5$~mJy listed in Table~\ref{tab:smg}.  According
to the models shown in Figure~4 of \citet{smail02}, we should expect
to find $\sim 0.25$ sources~arcmin$^{-2}$ to $S_{\rm 850\mu m}\sim
5$~mJy.  Neglecting cosmic variance, the nine observed radio-selected
SMGs to $S_{\rm 850\mu m}\sim 5$~mJy then imply an effective surface
area of $\sim 36$~arcmin$^{2}$.  The spectroscopic redshifts compiled
in Table~2 of \citet{chapman05} indicate that $\approx 44\%$ of the
radio-selected SMGs to $S_{\rm 850\mu m}\sim 5$~mJy lie at redshifts
outside the range $1.4<z<2.6$.  If we assume a Poisson distribution of
sources, then the total number of SMGs to $S_{\rm 850\mu m}\sim 5$~mJy
could be as high $9+\sqrt{9}=12$.  If we assume that the fraction of
interlopers among the $3$ unobserved objects is similar to the
fraction of interlopers among objects which are observed, then we
expect an upper limit of two SMGs with $S_{\rm 850\mu m}\sim 5$~mJy
that are unobserved and that lie between redshifts $1.4<z<2.6$.  If we
conservatively assume that these two sources are not selected by the
optical and/or near-IR criteria, and they have bolometric SFRs of
$\sim 1500$~$\sfr$ (similar to the average SFR found for the $5$
spectroscopically confirmed radio-selected SMGs in
Table~\ref{tab:smg}), then the inferred {\it additional} SFRD provided
by these two SMGs would be $\sim 3000$~$\sfr$ divided by the volume
subtended by $36$~arcmin$^{2}$ at redshifts $1.4<z<2.6$, or $\sim
0.022$~$\sfr$~Mpc$^{-3}$.  We note that this should be treated as an
upper limit for several reasons.  First, we have assumed the maximum
number of unobserved sources allowed by Poisson statistics.  Second,
we have assumed an interloper fraction among these unobserved sources
that is the same for the observed sources.  In general, one might
expect the contamination fraction to be higher among the general SMG
population to $S_{\rm 850\mu m}\sim 5$~mJy (where an accurate radio
position may not be known) than would be inferred from the
radio-selected SMG surveys.  Finally, we have assumed that all of the
unobserved sources cannot be selected by their optical and/or near-IR
colors.  Neglecting any overlap, radio-selected SMGs to $S_{\rm 850\mu
m}\sim 5$~mJy contribute $\sim 0.05$~$\sfr$~Mpc$^{-3}$ to the SFRD
between redshifts $1.4<z<2.6$.  However, our conservative calculation
indicates that radio-selected SMGs to $S_{\rm 850\mu m}\sim 5$~mJy
that are not selected by optical ($\ugr$) and/or near-IR ($\bzk$
and/or DRG) surveys make a small ($\la 0.022$~$\sfr$~Mpc$^{-3}$ or
$\la 15\%$) {\it additional} contribution to the SFRD between
redshifts $1.4<z<2.6$.

\section{Conclusions}

We have taken advantage of the extensive multi-wavelength data in the
GOODS-North field to select galaxies at $z\sim 2$ based on their
optical and near-IR colors and to compare them in a consistent manner.
Our own ground-based optical and near-IR images are used to select
galaxies based on their $U_{\rm n}G{\cal R}$, $\bzk$, and $\jmk$
colors.  Additional rest-frame UV spectroscopy for $25\%$ of optically
selected candidates allows us to quantify the redshift selection
functions for the various samples.  We use the deep {\it Chandra} 2~Ms
X-ray data to determine the influence of AGN and estimate bolometric
SFRs for galaxies in the optical and near-IR samples.  We also use the
deep {\it Spitzer}/IRAC data in the GOODS-North field in considering
the stellar populations and masses of galaxies selected in different
samples.  Our analysis employs the same multi-wavelength data for a
single field (GOODS-North), using the same photometric measurement
techniques, for a consistent comparison between galaxies selected by
their optical and near-IR colors.  Our main conclusions are as
follows:

1. Star-forming galaxies at $z\sim 2$ selected by their $U_{\rm
   n}G{\cal R}$ colors (i.e., $\ugr$ galaxies) and their $\bzk$ colors
   (i.e., $\bzk$/SF galaxies) have optical and near-IR color
   distributions that indicate significant overlap ($\sim 70-80\%$)
   between the two samples.  Photometric scatter could account for the
   colors of at least half of those galaxies missed by one set of
   criteria or the other.  The $\bzk$/SF criteria are less efficient
   in selecting (younger) $\ks>21$ galaxies at redshifts $1.4<z<2.6$,
   while the $\ugr$ criteria are less efficient in selecting near-IR
   bright (e.g., $\ks<20$) objects.  Distant red galaxies (DRGs;
   including both reddened star-forming and passively evolving
   galaxies) selected to have $\jmk>2.3$ show near-IR colors that are
   $1-1.5$ magnitudes redder than for samples of star-forming
   galaxies.  Criteria aimed at selecting passively evolving galaxies
   based on their $\bzk$ colors (i.e., $\bzk$/PE galaxies) by design
   have red near-IR colors, but we find that the redshift
   distributions of $\bzk$/PE galaxies and DRGs have very little
   overlap.

2. The deep X-ray data show that almost all of the directly detected
   X-ray sources in the samples have hard band emission and
   X-ray/optical flux ratios indicating they are likely AGN.  Much of
   this AGN contamination occurs for magnitudes $\ks<20$.  We identify
   $5$ objects that are detected in the soft band X-ray data and are
   likely star-forming galaxies based on their absence in the
   hard-band X-ray data, optical magnitudes $R>22$, and absence of
   obvious AGN features for those with spectra.  We stacked the X-ray
   data for all likely star-forming galaxies (i.e., those undetected
   in X-rays and the $5$ galaxies discussed above), excluding likely
   AGN.  The stacking analysis shows that the star formation rate
   (SFR) distributions of $\ugr$ and $\bzk$/SF galaxies and DRGs are
   very similar as a function of $\ks$ magnitude.  Galaxies with
   $\ks<20$ have average SFRs of $\sim 120$~$\sfr$, a factor of two to
   three higher than $\ks>20.5$ galaxies.  Previous studies point to a
   similarity in the metallicities, clustering, and stellar masses of
   $\ks<20$ optical and near-IR selected galaxies (e.g.,
   \citealt{shapley04}, \citealt{adelberger2005}).  In this work we
   show that the bolometric SFRs of optical and near-IR selected
   galaxies are also very similar when subjected to a common near-IR
   magnitude.

3. Near-IR selection of star forming galaxies should be more immune to
   the effects of dust obscuration than optical surveys.  However, the
   $\ugr$, $\bzk$/SF, and DRG samples show very similar SFRs as a
   function of near-IR color for galaxies with $\zmk<3$.  The SFRs
   inferred for $\bzk$/SF galaxies which are not optically-selected
   are very similar to $\bzk$/SF galaxies which do satisfy the optical
   criteria, suggesting that star-forming galaxies in near-IR samples
   that are missed by optical criteria do not harbor large numbers of
   heavily reddened galaxies.  Furthermore, the optical and $\bzk$/SF
   samples host an approximately equal number of submillimeter
   galaxies (SMGs).  Despite their very high SFRs ($\sim
   1000$~$\sfr$), SMGs missed by optical and near-IR selection have
   insufficient space densities to make a significant contribution
   ($\la 10\%$, see conclusion 6 below) to the total census of star
   formation at redshifts $1.4\la z\la 2.6$.
   
4. We identify a population of extremely red $\bzk$ and DRG galaxies
   with $\zmk\ga 3$.  The stacked X-ray data indicate these red
   galaxies have little, if any, current star formation.  The absence
   of X-ray emission from these objects also suggests that low
   luminosity AGN and low mass X-ray binaries contribute little X-ray
   emission in star-forming galaxies compared with the emission
   produced from more direct tracers of the current star formation
   rate, such as high mass X-ray binaries.  We further demonstrate the
   utility of deep X-ray data to constrain the stellar populations of
   these extremely red galaxies, and find that they must be described
   by declining star formation histories.  Almost all of these
   passively evolving galaxies satisfy the IERO criteria of
   \citet{yan04}.  We find that optical selection includes a subset of
   galaxies with stellar masses similar to those inferred for IEROs,
   but which are forming stars at a prodigious rate.  The stellar mass
   estimates from SED modeling (e.g., \citealt{yan04},
   \citealt{forster04}) and bolometric SFR estimates from the deep
   X-ray data (this work) suggest that the presence or absence of star
   formation may be the only significant difference between optical
   and near-IR selected massive galaxies ($M^*>10^{11}$~M$_{\odot}$),
   and the difference in SFR may be temporal.

5. We find evidence for a significant presence of passively evolving
   galaxies at redshifts $z\ga 2$ compared with their space density at
   lower redshifts, $1.4<z<2.0$.  Our analysis suggests that a single
   color technique using the $\zmk$ or $\jmk$ color allows for a more
   practical method selecting passively evolving galaxies with $z\ga
   2$ than the $\bzk$/PE criteria, as the latter would require
   excessively deep $B$-band data to accurately determine the space
   densities of passively evolving galaxies at $z\ga 2$.

6. Finally, we consider the contribution of optical and near-IR
   selected galaxies to the SFRD at $z\sim 2$, taking into account the
   overlap between the samples and their respective redshift
   distributions.  We find that $\ugr$ and $\bzk$/SF galaxies to
   $\ks=22$, and DRG galaxies to $\ks=21$, account for an SFRD of
   $\sim 0.10\pm0.02$~$\sfr$~Mpc$^{-3}$ between redshifts $1.4<z<2.6$.
   Approximately $87\%$ of this total comes from optically-selected
   galaxies to ${\cal R}=25.5$ and $\ks = 22$ and near-IR selected
   $\bzk$ galaxies to $\ks=22$, and $13\%$ from $\ks<21$ DRGs not
   selected by the $\ugr$ or $\bzk$ criteria.  Of the known
   radio-selected SMGs to $S_{\rm 850\mu m}\sim 4$~mJy in the GOODS-N
   field with redshifts $1.4<z<2.6$, $\ga 80\%$ could be selected by
   the $\ugr$, $\bzk$, and/or DRG criteria.

\acknowledgements

We thank Scott Chapman for discussions regarding submillimeter
galaxies in the GOODS-North field.  Haojing Yan kindly provided
stellar mass estimates for IEROs.  We thank David Alexander for his
suggestions regarding the use of the X-ray data, and Amy Barger for
her comments regarding Table~\ref{tab:smg}.  NAR, DKE, and CCS are
supported by grant AST 03-07263 from the National Science Foundation
and by the David and Lucile Packard Foundation.  AES and KLA are
supported by the Miller Foundation and the Carnegie Institute of
Washington, respectively.

\bibliographystyle{apj}

\end{document}